\documentclass{article}

\usepackage{PRIMEarxiv}

\usepackage[utf8]{inputenc} 
\usepackage[T1]{fontenc}    
\usepackage{hyperref}       
\usepackage{url}            
\usepackage{booktabs}       
\usepackage{nicefrac}       
\usepackage{microtype}      
\usepackage{lipsum}
\usepackage{fancyhdr}       
\usepackage{graphicx}       
\graphicspath{{media/}}     
\usepackage{amssymb,amsfonts}

\usepackage{threeparttable}         
\usepackage{multirow}
\usepackage{makecell}

\usepackage{adjustbox}
\usepackage{mathrsfs}

\usepackage{chngcntr}
\usepackage{subcaption}

\usepackage{hyperref}

\usepackage[centertags]{amsmath}			
\usepackage{stmaryrd}					
\usepackage{amsthm}					
\usepackage{newlfont}					
\usepackage{layouts}					
\usepackage{longtable,rotating}			
\title{DD\_R\lowercase{o}TIR: Dual-Domain Image Registration via Image Translation and Hierarchical Feature-matching
}

\author{
  Ruixiong Wang\thanks{\textbf{Corresponding author}: ORCID: 0000-0003-1824-185X}
  , Stephen Cross, ..., Alin Achim
  \\
  University of Bristol \\
  Bristol, United Kingdom\\
  \texttt{\{ruixiong.wang, stephen.cross, ..., alin.achim\}@bristol.ak,uk} \\
}

\newcounter{supfigure}
\renewcommand{\thesupfigure}{S\arabic{supfigure}}

\newcounter{supequation}
\renewcommand{\thesupequation}{S\arabic{supequation}}

\newcounter{supsection}
\renewcommand{\thesupsection}{S\arabic{supsection}}

\newcounter{supsubsection}[supsection]

\begin{document}
\maketitle

\begin{abstract}

Microscopy images obtained from multiple camera lenses or sensors in biological experiments offer a comprehensive understanding of the objects from diverse perspectives. However, using multiple microscope setups increases the risk of misalignment of identical target features across different modalities. Thus, multimodal image registration is crucial. In this work, we build upon previous successes in biological image translation (XAcGAN) and mono-modal image registration (RoTIR) to develop a deep learning model, Dual-Domain RoTIR (DD\_RoTIR), specifically designed to address these challenges. While GAN-based translation models are often considered inadequate for multimodal image registration, we enhance registration accuracy by employing a feature-matching algorithm based on Transformers and rotation equivariant networks. Additionally, hierarchical feature matching is utilized to tackle the complexities of multimodal image registration. Our results demonstrate that the DD\_RoTIR model exhibits strong applicability and robustness across multiple microscopy image datasets.

\end{abstract}

\keywords{Multimodal image registration \and microscopy imaging \and feature matching}

\section{Introduction}
\label{sec:intro}

Image registration for multimodal images poses a greater challenge compared to mono-modal registration since the semantic features used for similarity comparison and spatial alignment need to be extracted using inconsistent strategies. Automatic feature-extracting approaches encounter difficulties in maintaining consistent parallel extractions. Solutions to these challenges in deep learning-based approaches typically involve employing a multi-to-mono modality transformation method, unifying images from different modalities into a consistent modality, and then applying the mono-image registration methods. In addition to image translation, image fusion represents another approach to multimodal pre-processing. Image fusion aims to generate a single synthetic image that integrates the complementary advantages of images from multiple modalities, resulting in enhanced visual perception and information retrieval \cite{dey2019multi}. 

There exist strong interactions among image registration, translation and fusion. Successful outcomes by multimodal image registration can benefit image translation and fusion, as the registration results can be used as the foundations for translation and fusion. Conversely, excellent image translation and fusion approaches contribute to accurate multimodal registration, which implements the process of the multi-to-mono modality transformation. In practice, achieving registration with translation or fusion is not the ultimate goal; instead, these processes serve as fundamental steps, contributing to subsequent tasks such as image segmentation, object detection, target tracking and image interpretation \cite{xu2023murf}.

In the preceding work, We successfully tackled image registration of biological microscope images within the same domain \cite{wang2024rotir}. The primary challenge faced in that process was the absence of ground truth for supervised training and result evaluation. However, it is essential to note that the registration of mono-modal images is just a foundational step in biological image processing. The more intricate challenge lies in transcending the boundaries between image domains and facilitating alignment across images from multiple modalities. Encouragingly, the task of biological microscopy image translation, as presented in \cite{wang2023bright}, achieved promising results. This accomplishment now serves as a valuable tool for interpreting and exchanging feature information between different modalities.

In this chapter, We propose the Dual-Domain RoTIR (DD\_RoTIR) model designed for multimodal image registration. We employ the image dataset containing CHO cell images, which was used in the image translation task, to execute the registration model task. The objective image domains for this task include both brightfield and fluorescent images from the CHO cell image dataset, and we adhere to the same strategies for dataset synthesis as in the mono-modal image registration task. Recognizing the heightened complexities associated with multiple modalities, we enhance the deep-learning-based registration model by upgrading and emphasizing feature-point detection to address these challenges. Instead of generating a single matching map for each pair of images, the dual-domain image registration model produces two levels of matching maps from two feature fields with different resolutions. The integration of two matching maps effectively mitigates incorrect detections, thereby improving the overall accuracy of the model. This innovative approach has demonstrated its applicability to other datasets, establishing its versatility across multiple biological microscopy image datasets.

The paper is organized as follows: Section \ref{sec:background} provides an overview of the background knowledge that forms the foundation of our work. In Section \ref{chap5:sec01} to Section\ref{chap5:sec03}, we delve into the details of the algorithms employed, covering aspects such as network architecture, the training dataset, and the loss function. The registration results produced by RoTIR are discussed in Section \ref{chap5:sec04}. Finally, Section \ref{chap5:sec05} encapsulates and summarizes the outcomes of the study.

\section{Background}
\label{sec:background}

Registration of images from multiple modalities can be regarded as a special scenario of image registration. Except for the challenges for mono-modal image registration, a crucial task is to overcome the barriers between modalities and align contrasting features with different natures \cite{jiang2021review}.  

\subsection{Visible and Infrared Image Registration}
\label{chap2:sec03sub0401}

Visible and infrared images are commonly used as multimodal images in object tracking. Previous results have demonstrated that independently applying a single modality has significant limitations in certain circumstances. Therefore researchers turn to image fusion to improve object tracking results, necessitating registration for visible and infrared images is required \cite{zhang2020object}. 

Traditional methods have received a series of successes using area-based and feature-based methods. These methods typically involve extracting some common features such as edges or corners \cite{hrkac2007infrared, ma2015non, yu2019grayscale}. Wang \textit{et al.} proposed a two-stage Transformer network that utilized conditional GANs to obtain mapped images and a Transformer module to generate warped images. Their methods enabled the backpropagation of multi-spectral registration loss, enabling end-to-end training \cite{wang2018infrared}. Arar \textit{et al.} adopted a similar idea, employing an image translation model to bypass the cross-modality similarity measurement and subsequently using STNs for registration \cite{arar2020unsupervised}. Xu \textit{et al.} proposed the MURF model to address multimodal image registration and fusion as a combined issue. MURF contained three procedures: shared information extraction, which transformed the multimodal to mono-modal representation for shared information extraction and variance elimination; multi-scale coarse registration for global rigid parallax correction; and fine registration and fusion to rectify local non-rigid offsets. The fusion process and the registration complement each other, leading to higher accuracy for both \cite{xu2023murf}. 

In the visible and infrared image registration, significant semantic information is often more discernible compared to biological images. Features such as the edges and corners are normally easier to extract and distinguish, simplifying the translation approaches. However, in biological microscopy images, maintaining the semantic features poses a more challenging task.

\subsection{Medical Image Registration}
\label{chap2:sec03sub0402}

The field of medical imaging constitutes the largest community in multimodal image registration. The natures of medical images vary based on the different sensors or data types, encompassing X-ray, ultrasonic (US), computer tomography (CT), positron emission tomography (PET) and magnetic resonance imaging (MRI). In the clinic, leveraging the advantages of multiple types enhances the accuracy of final diagnosis and treatment. Consequently, image registration for medical images holds paramount significance and a substantial body of academic research emanates from this field.

Chen \textit{et al.} developed a framework featuring a two-channel channel registration algorithm to address mutual information challenges. The Proxy Registration of Cross-modality Images (PROXI) model was designed to synthesize the proxy images, filling in missing modalities for alignment purposes. This approach enabled traditional multi-channel deformable registration algorithms to perform the subsequent registration. The model was specifically applied to MRI images, focusing on the inter- and intra-subject registrations of T1-weighted and T2-weighted MR brain images. The results demonstrated significant improvement compared to single-channel registration methodologies \cite{chen2017proxy}.
The coarse-to-fine concept, similar to that in the MURF models, was also utilized in the registration of MRI images with five modalities, Huang \textit{et al.} proposed an end-to-end coarse-to-fine network that sequentially addressed affine and deformable transformations. The approach involved applying a dual consistency constraint and a prior knowledge-based loss was applied to enhance the registration process \cite{huang2021coarse}.
Cao \textit{et al.} introduced the cross-domain fusion registration network (CDFRegNet), an unsupervised approach designed for the registration of CT to cone-beam CT (CBCT) images. The model incorporated an edge-guidance attention module to eliminate disturbances by false edges resulting from artefacts and inconsistent structure in CBCT, emphasizing inherent gradient features. Additionally, a cross-domain attention module was employed to enhance the efficiency of mapping and fusing domain-specific features. The CDFRegNet model demonstrated its efficacy in CT to CBCT image registration \cite{cao2022cdfregnet}. Similarly, Song \textit{et al.} utilized the concept of cross-domain attention in their registration work, incorporating a contrastive learning-based pre-trained method for high-level feature extraction from the multiple modalities following cross-modal attention learning \cite{song2022cross}. 
Multimodal image registration extends beyond registration within individual MRI or CT images; it is also adapted for registration between different modalities.
Han \textit{et al.} introduced an unsupervised, deformable registration network designed for MR-CT registration, specifically for invasive neurosurgery. The proposed network comprised a synthesis network for image combination and a dual-channel deformation network for registration, ultimately producing a diffeomorphic deformation field. The synthesis network, implemented using a probabilistic CycleGAN, generated both synthetic images and associated uncertainty. The uncertainty calculation provided principled and spatially varying weights for the dual channels. The network components were trained synchronously which allowed the synthetically generated intermediate representation to guide the deformable registration process \cite{han2022deformable}. 

\subsection{Registration of Biological Images}
\label{chap2:sec03sub0403}

Biological microscopy image registration, for both mono- or multi-modal, has received less attention compared to registration among other medical imaging modalities, including MRI, CT, US or X-rays. One contributing factor is that some microscope setups use multiple camera lenses or sensors for image acquisition which maintains the spatial relationships between images from different modalities unchanged, such as in the case of brightfield and fluorescent images obtained simultaneously. However, there are still specific demands and challenges for biological microscopy image registration. For instance, techniques like Correlative Light and Electron Microscopy (CLEM), electron microscopy (EM) and fluorescent microscopy (FM) provide images with high resolution and molecular specificity, respectively. These modalities offer advantages in molecule identification and subcellular structure assignment. The previous approach, such as one by Hodgson \textit{et al.}, involved a conjunct process of cell centre detection and transformation parameter calculation. However, these methods faced difficulties when dealing with small cells, and the registration process failed without reliable detection information \cite{hodgson2014retracing}.

One significant contribution to biological image registration is the work by Pielaowski \textit{et al.}, who proposed the Contrastive Multimodal Image Representation for Registrations (CoMIRs) model and successfully applied it to histological images. In this approach, Pielaowski \textit{et al.} adopted the strategy of reducing multimodal to a single modality but employed a contrastive loss based on noise-contrastive estimation (InforNCE) to train neural networks for each modality \cite{oord2018representation}. Meanwhile, they also incorporated a hyperparameter-free modification InfoNCE to enforce the rotational equivariance of the learned representations. The contrast learning in CoMIRs learnt to minimize the difference between corresponding feature representations and maximize those from distant positions \cite{hu2019towards}. CoMIRs represented a novel approach as the first attempt to generate dense representations for image modalities with significant variance. The outputs from CoMIRs could be leveraged by existing mono-modal registration based on feature- or intensity-based registration methods. Unlike image translation and fusion approaches, CoMIRs transferred multimodal images by creating intermediate feature fields representing key features while maintaining the spatial locations. As a supervised model, CoMIRs, with a sophisticated training data augmentation scheme, demonstrated the feasibility of training with a limited dataset. Pielaowski \textit{et al.} \cite{pielawski2020comir} also compared the performance of CoMIRs to image-to-image translation approaches using GANs. Their findings indicated that GAN-based models were inadequate for registration. This conclusion will be discussed later. 

Further work that combined CoMIRs with Intensity and Spatial Information-based Deformable Image registration (INSPIRE) \cite{ofverstedt2023inspire} achieved improved results in various datasets like remote sensing images or cytological images. However, the combined approach showed limited improvement for a histological dataset containing brightfield and SHG images. This dataset will be included in my work for performance comparison.

\section{Dual-Domain Image Registration Network Design}
\label{chap5:sec01}

\begin{figure}[t!]
	\centering
	\includegraphics[width=0.9\textwidth]{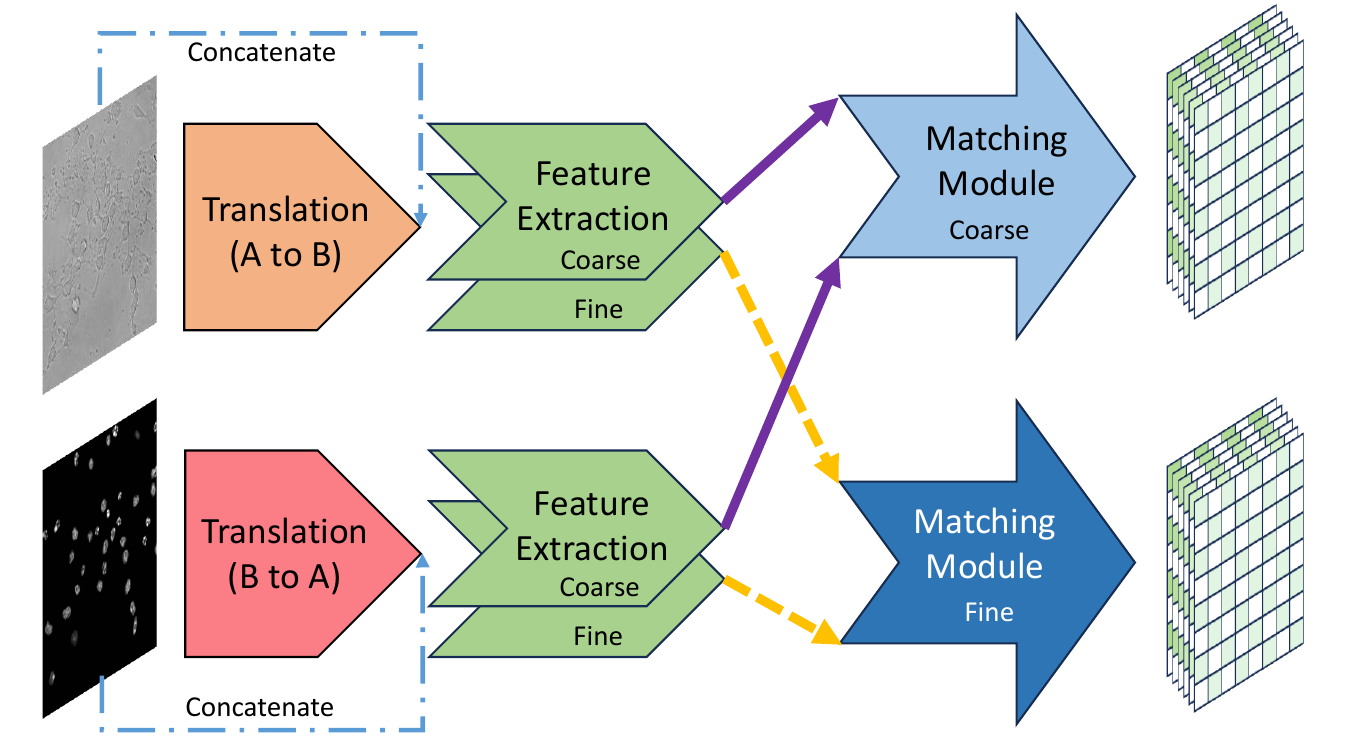}
    \caption{Workflow of the dual-domain image registration model. It comprises a pair of image translation modules, a pair of shared-weight feature extraction backbone networks, and a pair of matching modules for coarse and fine matching calculations.}
	\label{fig:workflow}
\end{figure}

The architecture of the dual-domain image registration network is illustrated in Figure \ref{fig:workflow}. Initially, a pair of image translation modules independently translate the input images to the other modalities and the translated images are then concatenated with the original inputs. These combined inputs are subsequently fed into the feature extraction backbones, which share the same parameters. The processed features from these backbones are sent to two hierarchies of matching modules, where matching maps are computed. Further details of each submodule are elaborated in the respective subsections.  

\subsection{Multi-to-Mono Modality Translation}
\label{chap5:sec01sub01}

In my work, I opted for image translation as the solution to bridge the gap between modalities. This choice allows for a relatively seamless transfer of the mono-modal image registration network initially designed for fish scale image registration. The decision to use image translation stems from the inherent challenges of training independent neural networks to transform images from different sources into a shared latent space. Moreover, in microscopy studies, there is often no clear guidance on how to merge multimodal images, as widely accepted prior knowledge is lacking. Creating an entirely new fused image domain for brightfield-fluorescent images would be complex and might lack practical significance. The contrastive learning of image representations has shown success for microscopy images \cite{nordling2023contrastive}, but I prefer retaining as much original information as possible to facilitate subsequent spatial alignment. 

Two image translation modules, each adhering to the architecture detailed in \cite{wang2023bright}, are employed for domain transition. Given that their primary objective is not the generation of highly accurate and authentic simulated images, the complexity of the architectures is streamlined to enhanced efficiency. The down-sampling levels are reduced from 5 to 4. Meanwhile, the mask guidance path for nuclei heath qualification is excluded. The training sources consist of brightfield and fluorescent images extracted from the central levels of the image stacks. Consequently, the numbers for input and output channels equal 1, aiming to balance the importance of each domain. The inputs for subsequent feature extraction are the concatenated of the original and corresponding synthetic images. The translation modules are pre-trained independently to provide a clear learning direction for the subsequent similarity comparison.

\subsection{Feature Extraction Backbone Networks}
\label{chap5:sec01sub02}

\begin{figure}[t!]
	\centering
	\includegraphics[width=0.8\textwidth]{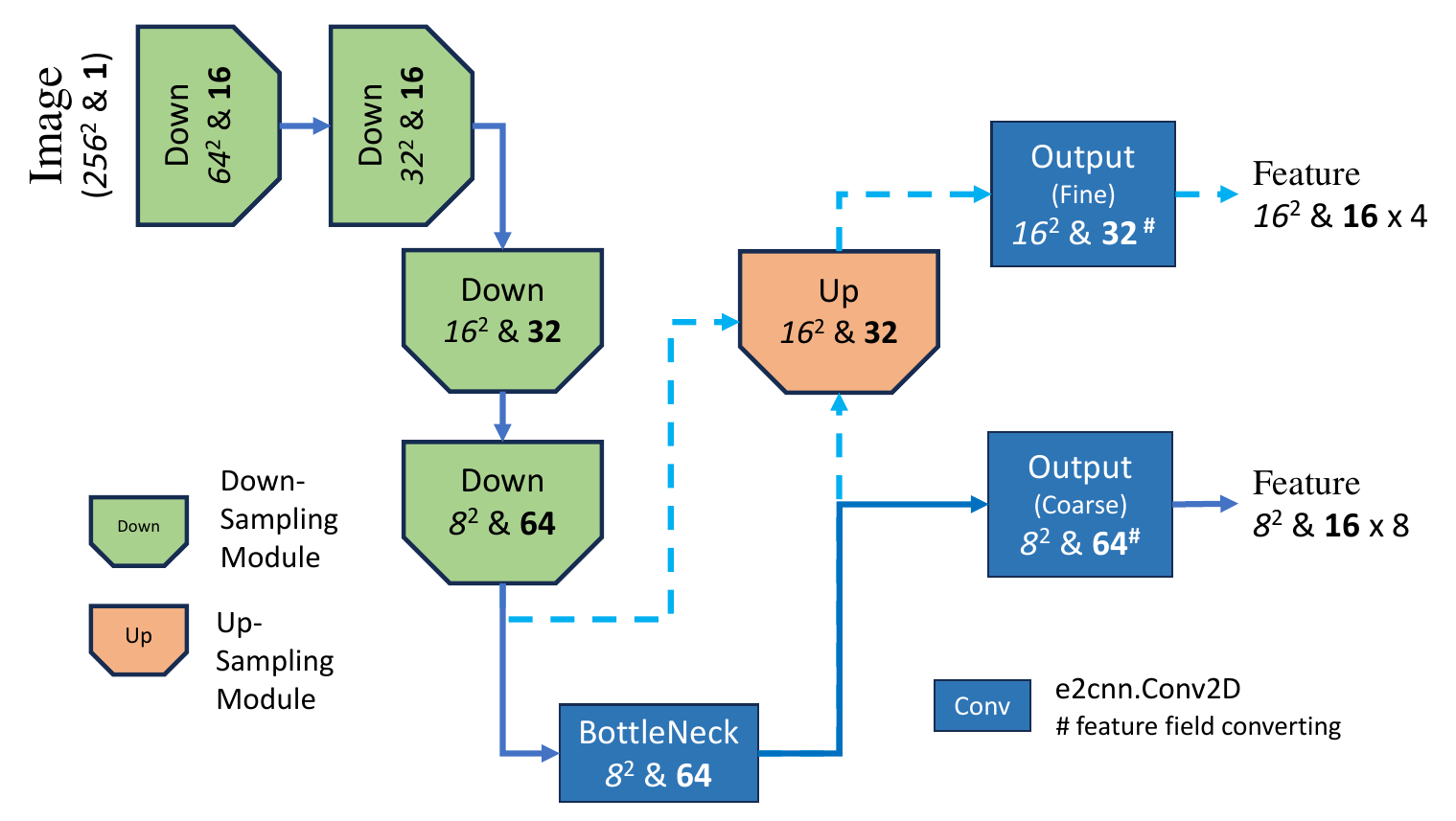}
    \caption{Architecture of feature extraction backbone network for dual-domain image registration. The structures of submodules for up-sampling and down-sampling are identical to those shown in figures form \cite{wang2024rotir}.}
	\label{fig:mmrbackbone}
\end{figure}

The backbone networks employed for feature extraction closely resemble those introduced in \cite{wang2023bright} for zebrafish scale image registration, with minor modifications. The down-sampling process, which reduces the inputs to vector fields with 1/16 of the original size, remains unchanged, except for the doubled input channel numbers due to the concatenation of inputs. However, the upsampling process undergoes revision. The modified model now incorporates only one up-sampling step, which transmits the features to the vector fields with irreducible representation with frequency 1 instead of regular representation. Consequently, the outputs of the up-sampling process are twice the size of those from the down-sampling process. Importantly, the outputs of the down-sampling and up-sampling process are not concatenated together. Instead, each set represents local features with different resolutions. These two sets of extracted features will be employed for hierarchical matching to achieve more precise key-point alignment. Figure \ref{fig:mmrbackbone} showcases the backbone network used for feature extraction in dual-domain image registration.

\subsection{Transformer-based Matching Modules and Hierarchical Feature Matching Algorithm}
\label{chap5:sec01sub03}

The matching modules of the DD\_RoTIR model are inspired by the key-point matching model LoFTR \cite{sun2021loftr} and share similar structures with the RoTIR model introduced \cite{wang2024rotir}. As depicted in Figure \ref{fig:workflow}, the workflow for each modality operates independently until reaching the matching modules. The translation and feature extraction modules independently process the individual resources and generate two sets of features with different hierarchies. The matching modules utilize the features with the same resolution from distinct resources and output groups of maps. Each group includes a confidence map of matching candidates along with supplementary information for affine transformation including rotation differences and coordinate refinements. 

One difference between the matching modules lies in the methods of confidence map regularization. The matching module for coarse features employs the dual-softmax operator \cite{cheng2021improving}, while the module for fine features utilizes the Sinkhorn iteration algorithms, consistent with the approach used in the RoTIR model. Additionally, there are slight differences in the linear projections of the output heads. Each matching module's linear projection is dedicated to a specific secondary exportation. 

During the training process, there is no interaction between the generation of coarse and fine maps. Losses are computed independently. The key distinction of the DD\_RoTIR model lies in the interaction between the two outputs from different hierarchies during evaluation and testing. Similar to the RoTIR model, an additional unlearning-based algorithm is utilized to produce the affine transformation parameters and construct the affine transformation matrices.

\subsubsection{Matching Map Calculation}
\label{chap5:sec01sub0301}

Pixel values on both confidence maps for feature matching represent pairs of matching patches from moving and fixed images. The coarse feature matching map reveals matching correlations of larger areas, it aims to set a loose restriction to the locations of matching pixels. On the contrary, the fine feature matching map provides more precise matching of local area. The methods for calculating confidence maps are different for them. As coarse matching refers to a larger area, the situation of one-to-multi correspondence commonly exists. However, the Sinkhorn iteration algorithm struggles to prove a distinct confidence score for matching candidates with multiple corresponding targets. I conducted an experiment to compare the performance of the dual-softmax operator and the Sinkhorn iteration algorithm, as detailed in Appendix \ref{App:sec03}. The results demonstrate that the dual-softmax operator exhibits advantages in handling one-to-multi feature point matching scenarios.

Nonetheless, the Sinkhorn iteration algorithm is still used for fine-matching map calculation. As in fine matching, the aim is to select the best correspondence avoiding the one-to-multi matching situation. Thus, the advantage of Sinkhorn iteration algorithms in one-to-one matching is revealed. Only the highest correspondence receives the highest score on the matching map, besides, the Sinkhorn algorithm has advantages in negative matching removing. Therefore, the dual-softmax operator and Sinkhorn iteration algorithms are applied respectively for coarse and fine matching.

\subsubsection{Rotation Detection and Coordinate Refinement}
\label{chap5:sec01sub0302}

Previous work on the RoTIR model demonstrated the feasibility of rotation detection and coordinate refinement in mono-modal image registration. The General E(2)-Equivariant Steerable CNNs are able to predict the angle between vectorized features \cite{e2cnn}. These processes are preserved in dual-domain registration. Scale variance detections are not considered in this work, as the images to be aligned in the task have the same resolution. During practical application, resolution factors could be obtained from the microscope settings. Rotation information and coordinate refinement can be calculated from both hierarchies of both linear projections. In constructing the network, I aimed to minimize duplication of effort and balance weights among loss components. To achieve this, I decided to extract one exportation from each hierarchy. Pixels on the coarse features represent wider areas of local image patches, making them more stable in angle detection. On the other hand, coordinate refinement is exploited solely on the fine feature matching maps as the location information on the maps is more precise.  

\section{Dataset Preparation}
\label{chap5:sec02}

\subsection{Image Pre-processing}
\label{chap5:sec02sub01}

In this study, the image dataset for multimodal image registration is applied to CHO-K1 cells, which were initially used for image translations and nuclei health quantification in \cite{wang2023bright}. The image acquisition and pre-processing procedures follow the same protocol outlined. Unlike the fish scale image dataset, a significant advantage is that all images in the CHO-K1 cell dataset are well-aligned, providing precise metrics for performance evaluation. The strategy for training dataset synthesis is consistent with that introduced in \cite{wang2023bright}, where aligned brightfield images and fluorescent images, along with the semantic segmentation masks, are utilised for generating ground-truth feature-matching maps. One difference compared to the dataset for image translation is that the input image size is maintained at $512\times512$px$^2$ without reshaping. 

\subsection{Ground-truth Matching Map Generation}
\label{chap5:sec02sub02}

The dataset synthesis concept involves employing randomly selected parameters to create the target transformation matrices and generate transformed image pairs as model inputs. The direct outputs from the DD\_RoTIR models consist of groups of maps. For efficient training loss calculation, pre-defined randomly selected parameters form two sets of ground truth maps for each resolution hierarchy. Figure \ref{fig:datasyth} illustrates the brief workflow of dataset synthesis and the integrated procedures for dataset establishment applied on a pair of brightfield and fluorescent images are presented below:

\begin{figure}[t!]
	\centering
	\includegraphics[width=0.75\textwidth]{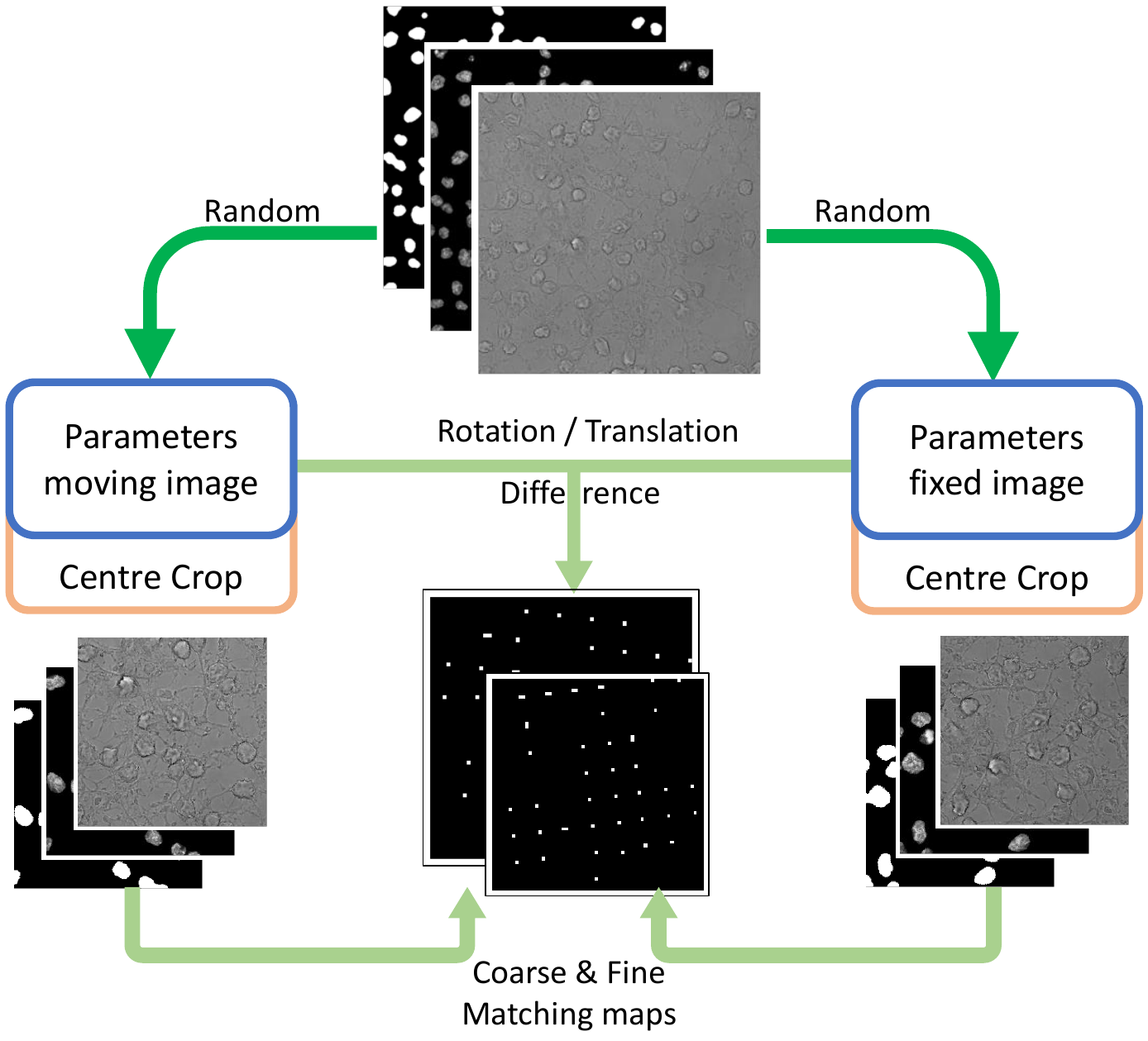}
    \caption{Dataset synthesis process for training and evaluation of the DD\_RoTIR Model.}
	\label{fig:datasyth}
\end{figure}

\begin{enumerate}
    \item A pre-processed image set consisting of aligned brightfield, fluorescent images and nucleus mask, all with sizes of $512\times512$px$^2$. The brightfield and fluorescent images are piecewise-linearly normalized to a range of $[-1,1]$.
    \item Randomly select parameters for the rotation angle $\theta$ and a pair of parameters for translation $t_x$ and $t_y$ for both dimensions from three independent continuous uniform distributions. \label{paraselect}
    \item Apply parameters $\theta$, $t_x$ and $t_y$ to determine the transformation function and crop the central part of the transformed images with a size of $256\times256$px$^2$ to form the moving image set.
    \item Repeat Step \ref{paraselect}, and form the fixed image set.
    \item Record the differences of the two randomly selected parameters and use them to create the affine transformation matrix.
    \item Apply an $8\times8$ grid on the masks of moving and fixed image sets. Calculate the mean values of patches in all grids, identifying patches with mean values higher than a threshold of 0 are candidates potentially corresponding to another image set. \label{coarsematch1}
    \item Transfer the affine image transformation matrix to the coordinate transformation matrix based on the Equation \ref{image2coord}. Find the matching pairs of candidates from two image sets. Create the coarse matching confidence map with a size of $64\times64$px$^2$, based on the coordinates of the matching patches on the $8\times8$ grid. \label{coarsematch2}
    \item Duplicate coarse matching map. Find the location of every positive pixel and insert the cosine and sine values. Then concatenate maps together to form a three-channel map.
    \item Repeat Step \ref{coarsematch1}, but with a $16\times16$ grid and threshold of 0.25 to identify the candidates.  
    \item Repeat Step \ref{coarsematch2} to generate fine matching confidence map with a size of $256\times256$px$^2$.
    \item Duplicate fine matching map. At the location of every positive pixel, insert the coordinate refinement values in both directions. Then concatenate maps together to form a three-channel map.    
\end{enumerate}

In the dataset, there are two pairs of brightfield and fluorescent images with sizes of $256\times256$px$^2$, and the cross-combination of images from the dual domains serves as inputs of the model. The remaining aligned images are not involved in the deep learning model training but are reserved for result exhibition. Given that they are cropped from double the size of the original image, there will be no padding issues as long as the translation distances are within a controlled range. Affine image transformation matrix and rotation angle are utilized for performance evaluation. The two sets of synthetic maps sized $64\times64$px$^2$ and $256\times256$px$^2$ respectively, are used in loss calculation. The datasets for training and performance testing remain consistent with those used in the XAcGAN model dataset.

\section{Methods}
\label{chap5:sec03}

\subsection{Loss Computation}
\label{chap5:sec03sub01}

The loss function for the dual-domain image registration network inherits the approach used in the RoTIR model. Losses for different hierarchies are calculated independently. The loss functions for the fine matching level and coordinate refinement remain the same as those in \cite{wang2024rotir}. The computation of the rotation angle difference is also consistent. 

However, with the dual-softmax operator being applied in the coarse matching level, there are slight changes to the loss function for the confidence map. 
Since the dual-softmax operator is used. Due
to the presence of one-to-multi matching, the confidence scores for positive matchings on coarse matching maps are not constant at $1$. Specifically, values for positive pixels on the confidence map generated by the dual-softmax operator vary within the range $(0,1]$, depending on the number of matching candidates. Equation \ref{eq:lossconfpos} and \ref{eq:lossconfneg} demonstrate the loss functions for the coarse matching: 

\begin{align}
    \mathcal{L}_{Coarse\_Conf}^{pos} &= \frac{1}{\left| M^{gt}\right|}\sum_{(i,j)\in M^{gt}} - \log\Big(1 + \big( M^{c}_{(i,j)} - M^{gt}_{(i,j)} \big)\Big)  - \log\Big(1 - \big( M^{c}_{(i,j)} - M^{gt}_{(i,j)}\big)\Big) \label{eq:lossconfpos}\\    
    \mathcal{L}_{Coarse\_Conf}^{neg} &= \frac{1}{\left| \overline{M^{gt}} \right|} \sum_{(i,j)\in \overline{M^{gt}}} - \log \Big( 1 - M^c_{(i,j)} \Big)  \label{eq:lossconfneg}     
\end{align}

\noindent 
$M^c$ represents the confidence map computed by the model. $M^{gt}$ and $\overline{M^{gt}}$ are the ground-truth positive and negative matching maps. The integrated loss of the confidence map is a weighted sum of positive and negative pixels. The weights for negative pixels are not constant; a higher weight is assigned to negative pixels that correspond to pairs of candidates from moving and fixed images but lack correspondence. This is accomplished by offering enhancing masks, serving the same function as the rectangular indicator boxes in RoTIR. For the fine-matching loss functions, enhancing masks are also applied. Differing from those in coarse matching, the enhancing masks are calculated from coarse matching maps. During the dataset synthesis, matching pixels on coarse matching maps indicate areas double the size of those on fine matching maps. As a result, pixels on fine-matching maps that refer to patches matched on the coarse-matching levels but not on the fine-matching levels receive higher weights during calculation. The confidence loss function for coarse and fine matching levels is shown in Equation \ref{eq:confloss}:

\begin{equation}\label{eq:confloss}
    \mathcal{L}_{Conf} = \mathcal{L}_{Conf}^{pos} + \lambda_{1} \cdot \mathcal{L}_{Conf}^{neg_1} + \lambda_{2} \cdot \mathcal{L}_{Conf}^{neg_2} \\
\end{equation}

The coarse- and fine-matching losses are weighted sums of confidence losses with angle difference or coordinate displacement, respectively, as presented in Equation \ref{eqdd:total}. Here, $\alpha$ and $\beta$ represent weights that balance each term. The final loss is the direct sum of losses for two hierarchies.


\begin{equation}\label{eqdd:total}
    \mathcal{L}_{Total} = \underbrace{\mathcal{L}_{Coarse\_Conf} + \alpha \cdot \mathcal{L}_{Angle}}_{Coarse} + \underbrace{\mathcal{L}_{Fine\_Conf} + \beta \cdot \mathcal{L}_{Trans}}_{Fine}
\end{equation}

\subsection{Affine Transformation Matrix Computation}
\label{chap5:sec03sub02}

The ultimate objective of the process is to determine the affine transformation matrices aligning the moving images with fixed images. Coarse and fine matching levels employ thresholds of 0.15 and 0.25, respectively, for selecting matching pixels. For each selected pixel at coarse matching levels, cosine and sine values are extracted from the corresponding positions and rotation angles are calculated. These angles are not directly applied to compute the final matrices, as the reasoning will be elucidated in the subsequent result evaluation section. The Z-score or standard deviation method is employed to identify detected angles considered outliers. By eliminating pixels with abnormal detected angles, the remaining matching maps are used to construct filters for fine-level matching maps.

The extracted feature fields at coarse levels refer to patches from the original images that are twice the size of those at fine levels. This implies that each pixel from coarse feature fields corresponds to 4($2\times2$) pixels on fine feature fields. Consequently, each pixel on coarse matching maps relates to 16 ($4\times4$) matching pairs of fine matching features, and also to 16 pixels on fine matching maps. Therefore, the initially obtained coarse-matching maps are transformed into filtration maps with the same sizes as fine-matching maps and 16 times the number of pixels. These filters are applied to the filtered fine-matching maps, and the resulting pixels are used for matrix calculation. The algorithm for matrix calculation is explained in Supporting Information \ref{chap3:sec02sub02}, with the only difference being that the pre-determined angles are not employed in forming the matrices. 

\subsection{Evaluation Metrics}
\label{chap5:sec03sub03}

The dataset used for performance evaluation is synthesized using the same methodology as the training dataset. Therefore, it incorporates the ground truth information for both the rotation difference and transformation matrix. Rotation and translation are two key elements in the rigid transformation of my model, and I employ angle difference and corner displacement as the evaluation criteria. The angle is derived from the affine transformation matrices, as outlined in Equation \ref{affinematrix} in Supporting Information \ref{chap3:sec02sub02}. Additionally, the displacement of the four corners is calculated using the Euclidean distance as shown in Equation \ref{eq:cornerdis}:

\begin{equation}\label{eq:cornerdis}
    DIS_{corner} = \frac{1}{4} \sum_{i=1}^4\big\Vert C_i^{gt} - C_i^{c} \big\Vert_2
\end{equation}

\noindent here, $C^{gt}$ and $C^c$ denote the locations of corners of the transformed moving images determined using ground-truth matrices and calculated matrices, respectively.

While evaluation metrics are occasionally interconnected with loss functions in certain supervised learning training scenarios, the dual-domain image registration task differs. Here, losses are computed from the intermediate outputs, and they may not fully unveil the comprehensive performance of the model. Notably, in this context, the impact of missing some positive matchings can be negligible, whereas a single erroneous matching could significantly compromise the entire prediction. 

Simultaneously, the success ratio, a metric used in the CoMIR model \cite{pielawski2020comir}, is introduced for performance evaluation. This metric calculates the percentage of cases in which the corner displacement falls below a specified threshold throughout the entire trial. Typically, thresholds of 1\% and 5\% of the image size length are employed. The success ratio provides valuable insights into the applicability and universality of the model.

\section{Results}
\label{chap5:sec04}

\subsection{Model Configuration and Training}
\label{chap5:sec04sub01}

Coding implementation remains consistent with previous works. In the training process, weights in loss functions of coarse and fine underwent several trials to determine the current optimal configuration as listed in Table \ref{tab:parameter}. To streamline the entire model, the configurations of the XAcGAN models for image translation are simplified compared to the model in \cite{wang2023bright}. Two groups of configurations of feature extractions are applied, one employs cyclic groups of 4 ($C_4$) with 64 hidden channels, and the other one utilizes cyclic groups of 8 ($C_8$) with 128 hidden channels. The transformer-based matching module has the same configuration as the RoTIR model. The model is trained with an Adam optimizer. The iteration is set at $3\times10^{4}$, with the learning rate starting at 0.001 and halving every 5000 epochs.  

\begin{table}[htbp]
\centering
\caption{Parameters for Loss Function}
\setlength{\tabcolsep}{4mm}{
\begin{tabular}{ccccc}
\toprule
       & $\lambda_1$ & $\lambda_2$ & $\alpha$ & $\beta$ \\
\midrule
Coarse & 500         & 100         & 20       & -\\
Fine   & 50          & 10          & -        & 20 \\
\bottomrule
\end{tabular}}
\label{tab:parameter}
\end{table}

\subsection{Results on CHO Dataset}
\label{chap5:sec04sub02}

Rotation angles exert a substantial influence on image registration, and previous models have exhibited poor performance within a broader rotation range. Therefore, in the evaluation phase, testing datasets are categorized as "small" and "large" based on the rotations. Specifically, they are divided into groups with ground-truth absolute rotation angle within $(0^{\circ},45^{\circ}]$ and $(45^{\circ}, 90^{\circ}]$, respectively. 

To assess the improvement introduced by the upgraded DD\_RoTIR models to the mono-domain RoTIR models, the results from RoTIR models trained with datasets with different image modalities are involved. The reference models consist of two RoTIR models trained exclusively with either brightfield images or fluorescent images, and one RoTIR model trained with pairs of concatenated images of real and pseudo images (real brightfield and pseudo fluorescent images or pseudo brightfield and real fluorescent images). The pseudo images are synthesized by the same twain models utilized in DD\_RoTIR models, translating brightfield and fluorescent images bilaterally. For the testing dataset, 720 pairs of unaligned images are generated, with brightfield images serving as moving images and fluorescent images as fixed images. It is noteworthy that the rotation angle detection method within the RoTIR models in this experiment differs from those in the image registration works of zebrafish scales. Here, the rotation angles calculated by the output maps are disregarded, and coordinates of detected feature points are used for final rotation and translation detection. Further analysis will be presented in the following context. Table \ref{tab:ddrotirresult} provides the results of the dual domain image registration along with the reference models.

\begin{table}[htbp]
    \centering
    \caption{Experiment Results}
    \begin{threeparttable}
    \setlength{\tabcolsep}{4mm}{
    \begin{tabular}{c c c c c c c }
    \toprule
    \makecell[c]{Model} & \makecell[c]{Rotation\\ Range}\tnote{1} & \makecell{Angle \\ Error} & \makecell{Corner \\ Error}\tnote{2} & \makecell{Succ.\\ < 1\%} & \makecell{Succ.\\ < 5\%} & \makecell{Accs. \\ Angle}\tnote{3} \\
    \midrule   
    \multirow{2}*{\makecell[c]{RoTIR\\ Brightfield}} & $(\ \ 0^{\circ}, 45^{\circ}]$   &  0.31$^{\circ}$ &  1.20 & 92.2\% & 99.7\% &  2.93$^{\circ}$ \\    
    \cline{2-7}
                                                     & $(45^{\circ}, 90^{\circ}]$   &  0.34$^{\circ}$ &  1.28 & 93.2\% & 99.7\% &  2.83$^{\circ}$ \\
    \midrule
    \multirow{2}*{\makecell[c]{RoTIR\\ Fluorescent}} & $(\ \ 0^{\circ}, 45^{\circ}]$   &  5.32$^{\circ}$ & 16.99 & 32.8\% & 74.2\% &  5.77$^{\circ}$ \\
    \cline{2-7}
                                                     & $(45^{\circ}, 90^{\circ}]$  & 13.43$^{\circ}$ & 41.18 & 31.2\% & 63.6\% &  7.30$^{\circ}$ \\
    \midrule
    \multirow{2}*{\makecell[c]{RoTIR\\Concatenate}}  & $(\ \ 0^{\circ}, 45^{\circ}]$   &  2.36$^{\circ}$ &  8.21 & 58.2\% & 90.7\% &  3.93$^{\circ}$ \\
    \cline{2-7}
                                                     & $(45^{\circ}, 90^{\circ}]$   & 25.99$^{\circ}$ & 73.57 & 26.1\% & 53.1\% & 23.61$^{\circ}$ \\
    \midrule
    \multirow{2}*{\makecell[c]{DD\_RoTIR\\(4,64)}}   & $(\ \ 0^{\circ}, 45^{\circ}]$   &  0.94$^{\circ}$ &  3.07 & 79.6\% & 98.6\% &  6.61$^{\circ}$ \\
    \cline{2-7}
                                                     & $(45^{\circ}, 90^{\circ}]$   &  1.10$^{\circ}$ &  3.69 & 78.6\% & 97.6\% &  4.84$^{\circ}$ \\
    \midrule
    \multirow{2}*{\makecell[c]{DD\_RoTIR\\(8,128)}}  & $(\ \ 0^{\circ}, 45^{\circ}]$   &  0.44$^{\circ}$ &  1.71 & 85.0\% & 99.7\% &  2.02$^{\circ}$ \\
    \cline{2-7}
                                                     & $(45^{\circ}, 90^{\circ}]$   &  0.50$^{\circ}$ &  1.89 & 83.5\% & 99.3\% &  3.38$^{\circ}$ \\    
    \bottomrule
    \end{tabular}} 
    \begin{tablenotes}
        \footnotesize
        \item[*] For angle and corner errors, lower values indicate better performance. For success rates, higher values indicate better performance. 
        \item[1] Absolute rotation angle ranges.
        \item[2] Corner displacement calculated by Equation \ref{eq:cornerdis}.
        \item[3] Accessory rotation angles directly predicted by models.
    \end{tablenotes}
    \end{threeparttable}
    \label{tab:ddrotirresult}
\end{table}

\subsubsection{Evaluation of RoTIR Models}
\label{chap5:sec04sub0201}

\begin{figure}[t!]
	\centering
	\includegraphics[height=0.5\textheight]{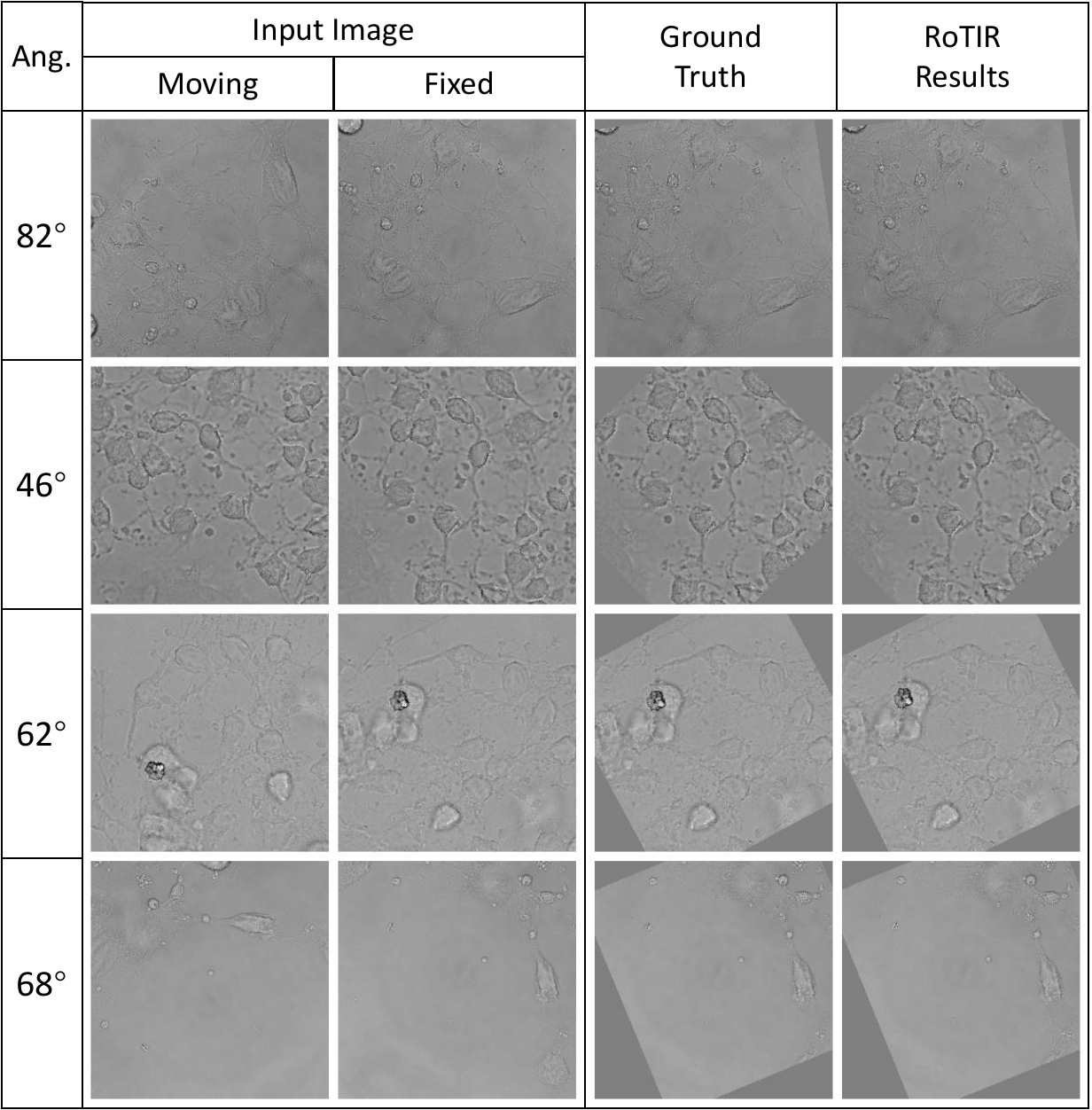}
    \caption{Mono-modal image registration result of the RoTIR model applied on brightfield images from CHE dataset. "Ang." represents the rotation angles between the moving and fixed images.}
	\label{fig:rotir_bright}
\end{figure}

The RoTIR model trained with the mono-modal of brightfield images demonstrated the best overall performance, achieving superior scores across all comparison metrics. Figure \ref{fig:rotir_bright} illustrates four examples of mono-modal image registration by the RoTIR model on brightfield images from the CHO dataset, where the rotation difference between the moving and fixed image are all larger than $45^{\circ}$. This outcome is consistent with the exceptional results previously observed in the image registration experiment of zebrafish scales. Examining Figure \ref{fig:rotir_bright},  it is apparent that brightfield images contain rich information evenly distributed throughout, allowing the RoTIR model to identify more matching points on the confidence matching maps. 

On the contrary, fluorescent images exhibit larger areas of blank backgrounds, lacking distinguishing features for the deep-learning model to learn. Besides, the nuclei in the fluorescent images display minimal textural variances, with only slight variations in edge shapes providing concealed information. Consequently, the RoTIR model encounters substantial challenges, resulting in the poorest performance among all the reference models.

The employment of RoTIR trained on concatenated images yields slightly improved outcomes compared to RoTIR trained solely on fluorescent images when rotation angles are within the absolute rotation range $[05^{\circ}, 45^{\circ}]$. However, the performance for precise prediction falls short of expectations. Furthermore, the results exhibit a significant decline in performance on large rotation angles with an average error of $25.99^{\circ}$ for rotation detection, a deviation far from anticipation. Therefore, it is evident that the conjunction of RoTIR and XAcGAN models holds potential for addressing the challenge, but a mere combination of these two models proves insufficient for achieving precise dual-domain image registration. Enhanced analysis of local features is necessary, and additional efforts are required to eliminate error-matching to identify true positive matches for registration calculations accurately. 

\subsubsection{Registration using DD\_RoTIR Models}
\label{chap5:sec04sub0202}

\begin{figure}[p]
	\centering
	\includegraphics[width=0.95\textwidth]{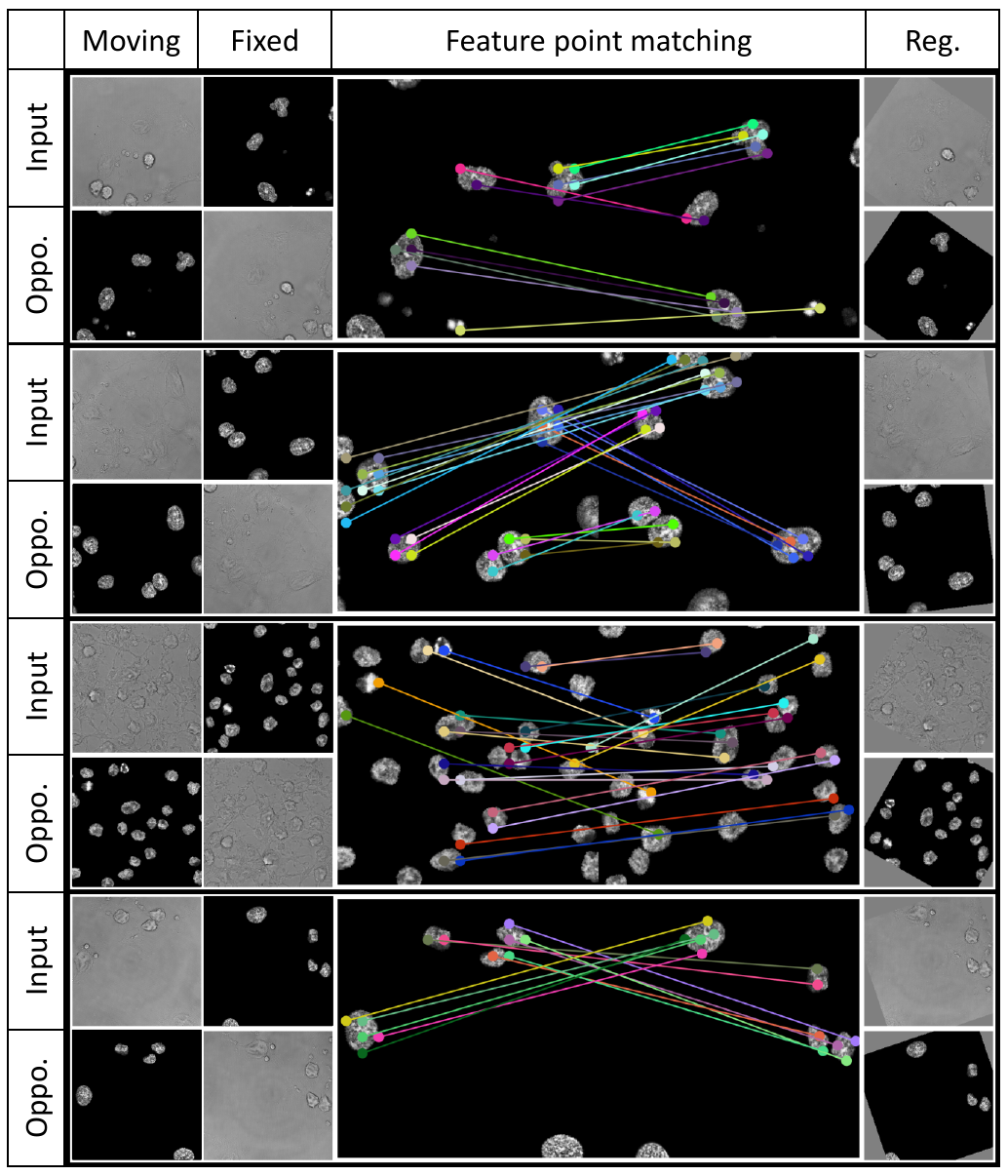}
    \caption{Multimodal image registration on brightfield and fluorescent images by DD\_RoTIR model (backbone configuration of (8, 128). Row "Oppo." represents the corresponding images on the opposite modalities. Column "Reg." shows the transformation result of moving images with corresponding fluorescent images.}
	\label{fig:ddregmatch}
\end{figure}

The results shown in Table \ref{tab:ddrotirresult} exhibit the promising performance of both DD\_RoTIR models in comparison to the RoTIR model. As expected, the model with a more complex feature extraction background network, using a Cyclic group of $C_8$ and 128 hidden channels, achieves better outcomes evaluated across all evaluation metrics. Notably, the mean values of angle difference and corner displacement for the DD\_RoTIR model with the best performance are approximate to those in the RoTIR model trained solely on brightfield images. This achievement is particularly remarkable considering the input images are sourced from multiple modalities. Figure \ref{fig:ddregmatch} illustrates four example results by the DD\_RoTIR model of $C_8$ and 128 hidden channels. In the figure, corresponding images from the opposite modalities are juxtaposed with the moving and fixed images. Transformation matrices are applied to the moving brightfield images and their corresponding fluorescent images. For clarity in presentation, the feature point matching visuals utilize the corresponding fluorescent images of the moving images and the input fixed images. This presentation approach is retained in the subsequent section for feature point matching visualization. 

The escalation of rotation differences between moving and fixed images poses heightened challenges for registration. Both the average values of angle differences and corner displacements increase while the success ratio decreases. This observed trend is also evident in the tests conducted exclusively on brightfield images. However, compared to the RoTIR model trained with concatenated images, the variances in every evaluation metric are relatively inconspicuous. This substantiates the claim that DD\_RoTIR models exhibit robustness in the face of varying rotation angles. The success ratio for corner displacement of less than 5\% remains close to 100\%, mirroring the performance of the best reference RoTIR model trained on the brightfield image dataset. The only remaining area for promotion lies in the success ratio of corner displacement of less than 1\%. 

\subsubsection{Translation Module Outputs}
\label{chap5:sec04sub0203}

\begin{figure}[t!]
	\centering
	\includegraphics[width=0.95\textwidth]{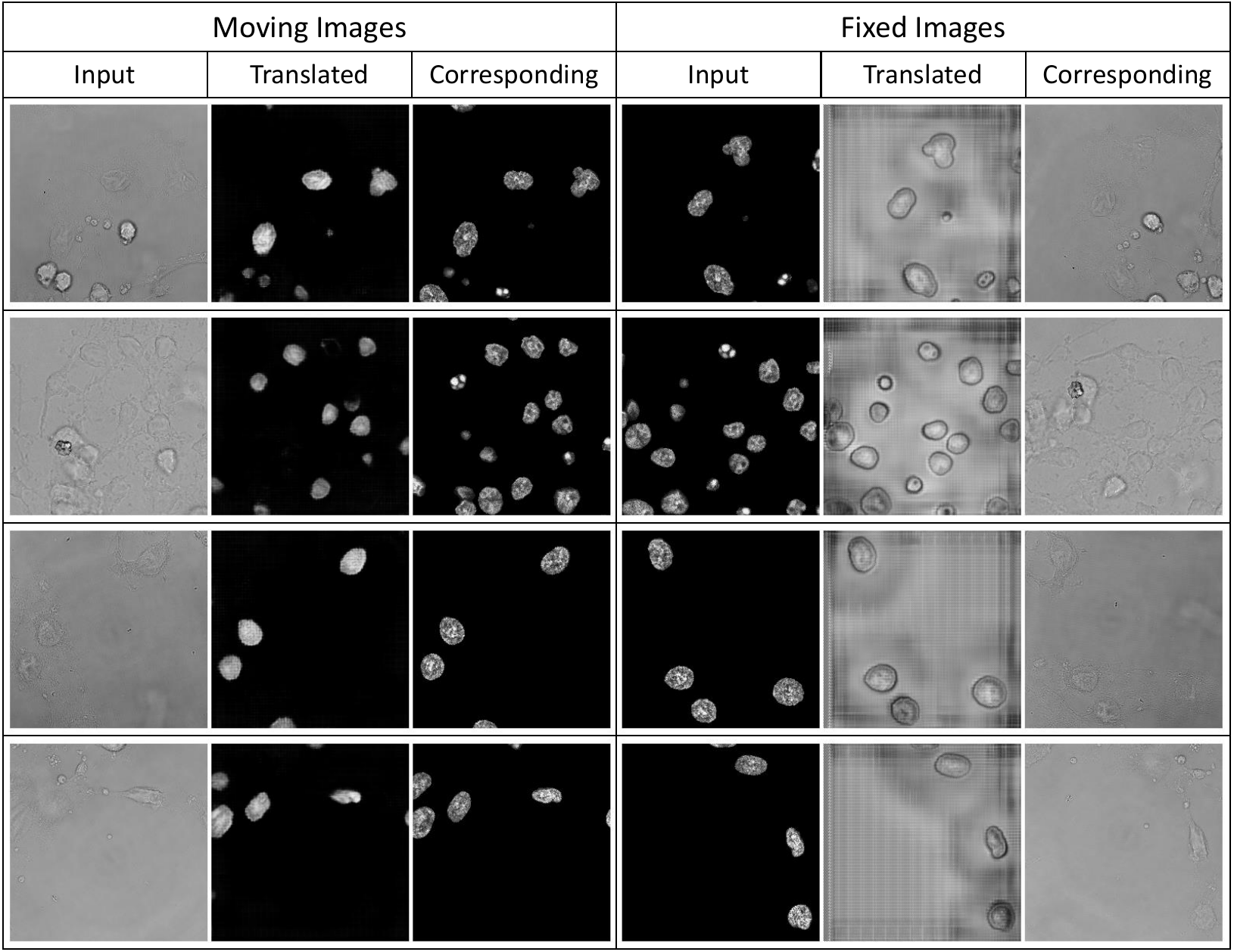}
    \caption{The pseudo images generated by the translation module of the DD\_RoTIR model. These pseudo images are translated from the input images and concatenated with the initial inputs for feature extraction.}
	\label{fig:trans_image}
\end{figure}

Nevertheless, the challenges inherent in bridging the gap between modalities always persist. The translated pseudo images generated by the translation modules within the DD\_RoTIR model are shown in Figure \ref{fig:trans_image}. In comparison to the translation result presented by XAcGAN model, there remains a discernible gap for the lightweight translation models. Details within the translated images exhibit blurriness, and the pseudo-brightfield images manifest a noticeable amount of noise. 
However, the upgrade of translation modules is constrained by the dataset size and the model efficiency. Fortunately, the feature extraction backbone networks, based on the Siamese Network approaches, prove effective in extracting significative features for similarity comparison from concatenated inputs, even when part of the inputs is pseudo. Coupled with the hierarchical matching algorithm, this compensates for the underperformance of translation modules.  

\subsubsection{Coarse Matching with Rotation Prediction}
\label{chap5:sec04sub0204}

\begin{figure}[t!]
	\centering
	\includegraphics[width=0.85\textwidth]{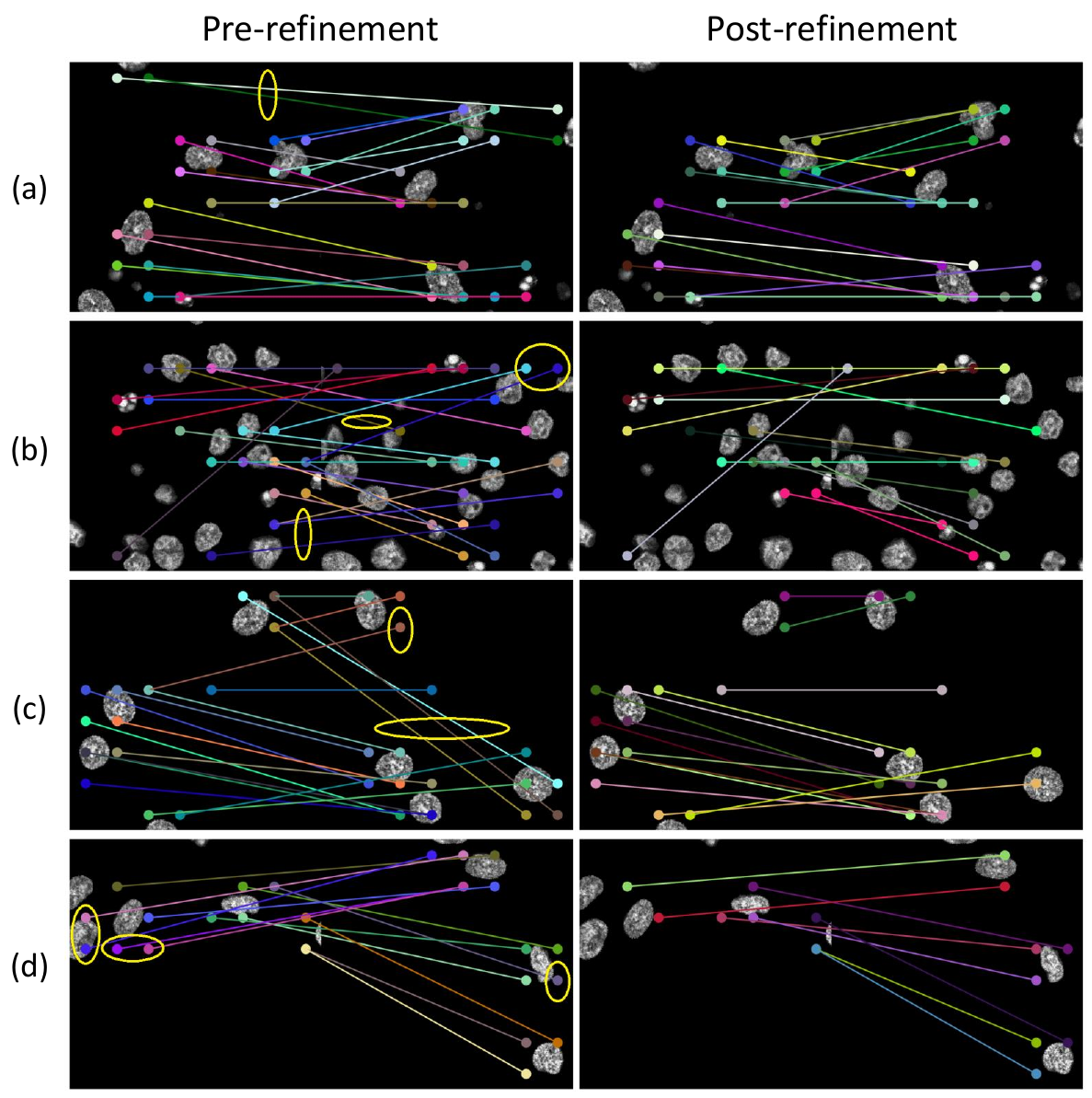}
    \caption{Refinement of coarse matching using rotation detection results. Yellow circles indicate matchings that are removed based on the filtration shown in Figure \ref{fig:coare_refine_angle}.}
	\label{fig:coare_refine}
\end{figure}

\begin{figure}[t!]
\centering
\begin{minipage}{0.45\textwidth}
    \centering
    \includegraphics[width=\linewidth]{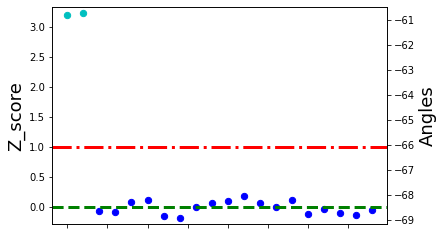}
\end{minipage}
\hspace{0.05\textwidth}
\begin{minipage}{0.45\textwidth}
    \centering
    \includegraphics[width=\linewidth]{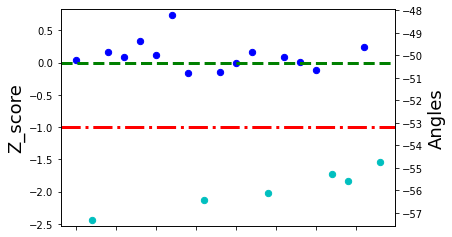}
\end{minipage}
\vspace{0.1cm}
\vspace{\baselineskip}
\begin{minipage}{0.45\textwidth}
    \centering
    \includegraphics[width=\linewidth]{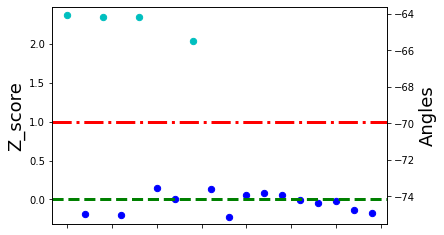}
\end{minipage}
\hspace{0.05\textwidth}
\begin{minipage}{0.45\textwidth}
    \centering
    \includegraphics[width=\linewidth]{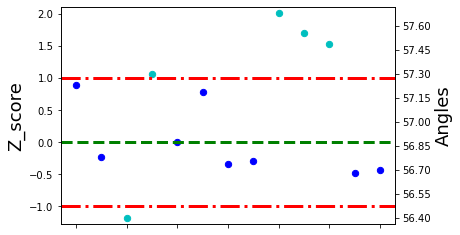}
\end{minipage}
\caption{Z-score filtration for coarse matching refinement. Matching points on coarse matching maps whose rotation angles exceed the limitation are removed. The green lines represent the median value of the detected angle, while the red lines denote the boundaries. Dark blue points represent items that are retained, while light blue points represent items that are removed.}
\label{fig:coare_refine_angle}
\end{figure}

From Table \ref{tab:ddrotirresult}, it is evident that the rotation angle detection of RoTIR models trained with different mono-modal datasets, as shown in the column of accessory rotation angles, is inferior to those calculated using coordinates of feature points. In RoTIR trained with brightfield images, the average rotation errors computed from the output angle maps are no higher than $3^{\circ}$ for both datasets of small and large rotations. While such results are generally acceptable, they are incomparable to those calculated solely by detected matching points, which are less than $0.35^{\circ}$. A similar phenomenon exists on concatenated images with the small-rotation dataset. Other results vary; however, since the predictions are significantly deviated from the ground truths, they are not considered for analysis. 

The reasons may come from multiple aspects. Unlike fish scales that feature distinguishable stripes, the local textures of brightfield images are more complex and lack clear directionality, making it challenging for the feature extraction backbone networks to predict precise directions. This presents an inherent disadvantage. Meanwhile, the total loss consists of multiple terms, with the most crucial being the confidence maps for correspondence computation. Increasing the weight for angle detection can influence the accuracy of matching map generation. On the other hand, if the output confidence maps have high reliability and provide more matching points, rotation accuracy will increase simultaneously. Acknowledging such a phenomenon, rotation detection by the RoTIR and DD\_RoTIR model is not used for final affine transformation matrix calculation. 

However, the results of the rotation detection are not discarded. Upon more in-depth analysis of the detected rotation angles, they are utilized for the refinement of the coarse matching. The Z-score calculation is employed to eliminate the false positive matches. Figure \ref{fig:coare_refine} displays four examples of coarse matching results. The results before refinement are listed in the left column. Figure \ref{fig:coare_refine_angle} illustrates the corresponding Z-score filtration process, where rotation angles that are evidently different are filtered out. The right column in Figure \ref{fig:coare_refine} exhibits the matching after filtration. The yellow circles on the pre-refined matching figures present the eliminated matches. In practice, even though the selected angles have relatively distinct differences from the ground truth, severe errors are removed ensuring the accuracy. 


\subsubsection{Fine Matching Results}
\label{chap5:sec04sub0205}

\begin{figure}[t!]
	\centering
	\includegraphics[width=0.85\textwidth]{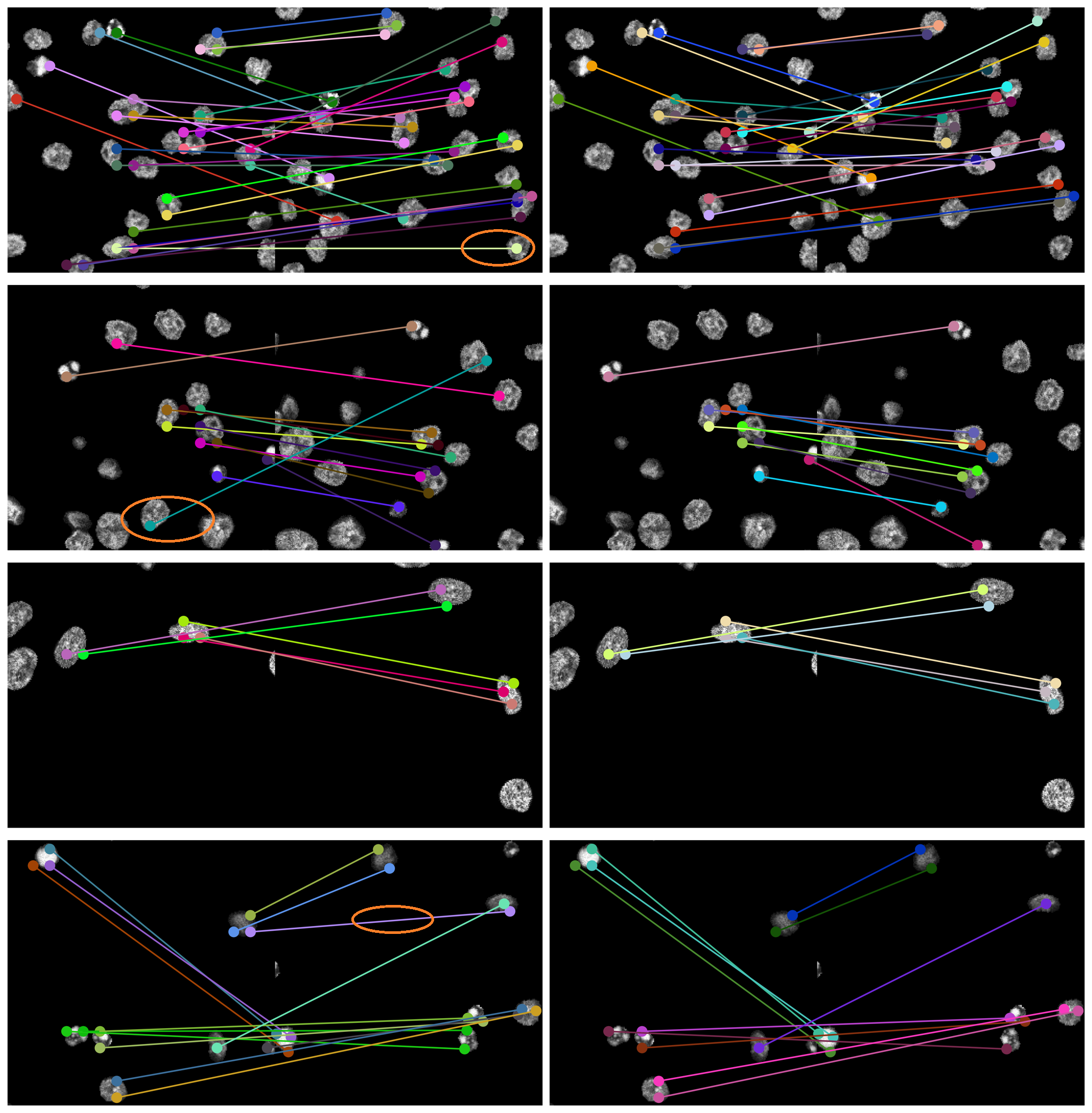}
    \caption{Final matching results. The right column shows the initial results directly from the fine matching results, orange circles refer to some cases in which the matching needs to be eliminated.}
	\label{fig:fine_refine}
\end{figure}

The DD\_RoTIR models were originally inspired by the feature point matching mechanism. Similar to the fish scale registration task mentioned in \cite{wang2024rotir}, DD\_RoTIR models do not require specific feature point determination. Feature extractions span the entire areas of the moving and fixed images to identify all the matching features. Feature points on moving images are situated at the centroids of the local image patches, with each patch having a size of $1/16 \times 1/16$ of the whole image. On the other hand, coordinate corrections are applied to the detected matching points on fixed images. The coarse matching results serve as the restriction, ensuring that fine-matching results must be located within the coarse matching. Figure \ref{fig:fine_refine} illustrates the comparison between initial fine-matching results (right column) and refined fine-matching results (left column). 

\subsection{Results on SHG Dataset}
\label{chap5:sec04sub03}

To assess the universality of the DD\_RoTIR model on other biological microscopy image datasets, a histological dataset that contains pairs of brightfield tissue micro-array and SHG images is introduced. The images in the dataset depict tissue section slides stained using traditional Hematoxylin-Eosin (H\&E), originally acquired for breast, pancreatic and kidney cancer studies. The brightfield images were captured using an Aperio CS2 Digital Pathology Scanner (Leica Biosystems, Wetzlar, Germany) at 40x magnification. The images in the dataset were subsequently processed and manually aligned for the multimodal image registration experiment \cite{pielawski2020comir, lu2022multimodal}. 

\subsubsection{Results using CoMIR Models}
\label{chap5:sec04sub0301}

Pielawski \textit{et al.} introduced the CoMIR models for multimodal image registration \cite{pielawski2020comir}. This approach transforms a pair of brightfield and SHG images into a hidden domain. Unlike image fusion, the images in the hidden domain do not require special biological significance but serve to provide features for image registration calculation. In their works, the CoMIR models were trained using critics based on Mean Squared Error (MSE) and cosine similarity. These models employed intensity- and feature-based image registration methods, namely $\alpha$-AMD (AMD for Asymmetric average minimal distance) and SIFT, to perform the registration of the transformed images. Meanwhile, they explored translation models, including CycleGAN \cite{zhu2017unpaired}, Pix2Pix \cite{Isola_2017_CVPR} and DRIT (Disentangled Representation for
Image-to-Image Translation) \cite{lee2018diverse}. Baseline performance was established using Mattes Mutual Information (Mattes MI) algorithms \cite{mattes2001nonrigid} with a (1+1) evolutionary strategy \cite{styner2000parametric}. 

In the CoMIR\_SIFT testing, 134 pairs of images with sizes of $834\times834$px$^2$ were used as inputs. The rotation was constrained within the absolute rotation range $[0^{\circ}, 30^{\circ}]$ and translations were up to $\pm100$px ($\pm12.0$\% of the image size) on both dimensions. The 134 pairs of images were evenly categorized into three groups based on the corner displacements between moving and fixed images. The "small" transformation referred to the displacement of less than 100px, similarly, "medium" and "large" indicated to within $(100, 200]$px and exceeding 200px, respectively.

The outcomes of CoMIR, as presented by Pielawski \textit{et al.}, are summarized in Table \ref{tab:comirresult}. Their findings indicated that GAN-based image generation methods struggled in the context of multimodal image registration. However, the CoMIR models achieved noteworthy results on histological data. Among the different approaches, CoMIR with SIFT demonstrated the most promising results and will serve as the benchmark for further comparison. 

\begin{table}[htbp]
    \centering
    \caption{CoMIR Results}
    \begin{threeparttable}
        \begin{tabular}{| p{5em}<{\centering} | p{3em}<{\centering} | p{5em}<{\centering} | p{5em}<{\centering} | p{5em}<{\centering} | p{5em}<{\centering} |}
        \hline
        Model & Metric & \makecell[c]{MS\tnote{1}\\ $\alpha$-AMD}  & $\alpha$-AMD & SIFT & Mattes MI\\
        \hline\hline
        \multirow{2}*{\makecell[c]{CoMIR\\MSE}} & < 1\% & 53.7\% & 36.6\% & \textbf{56.0\%} & 6.7\% \\ 
        \cline{2-6}
         & < 5\% & 67.2\% & 47.0\% & \textbf{73.1\%} & 22.4\%  \\
        \hline
        \multirow{2}*{\makecell[c]{CoMIR\\Cos.\tnote{2}}} & < 1\% & 40.3\% & 14.2\% & 38.8\% & 9.0\% \\ 
        \cline{2-6}
         & < 5\% & 46.3\% & 25.4\% & 51.5\% & 21.6\%  \\  
        \hline\hline
        \multirow{2}*{CycleGAN} & < 1\% & - & - & 0\% & 4.5\% \\ 
        \cline{2-6}
         & < 5\% & - & - & 0\% & 22.4\%  \\  
        \hline
        \multirow{2}*{Pix2Pix} & < 1\% & - & - & 0\% & 5.2\% \\ 
        \cline{2-6}
         & < 5\% & - & - & 0\% & 22.4\%  \\  
        \hline
        \multirow{2}*{DRIT} & < 1\% & - & - & 0\% & 4.5\% \\ 
        \cline{2-6}
         & < 5\% & - & - & 0\% & 23.1\%  \\  
        \hline
        \end{tabular}
    \begin{tablenotes}
        \footnotesize
        \item[1] Multiple scale $\alpha$-AMD.
        \item[2] CoMIR with cosine similarity.
    \end{tablenotes}
    \end{threeparttable}
    \label{tab:comirresult}
\end{table}

\subsubsection{Dataset Processing}
\label{chap5:sec04sub0302}

To align the brightfield and SHG images with the DD\_RoTIR model, a few adjustments were made in preparation for testing, ensuring consistency with the model configurations and the training details previously outlined for the DD\_RoTIR model trained on the CHO dataset presented, as presented in Section \ref{chap5:sec04sub01}. These modifications included channel compression and image cropping. Firstly, consideration was given to the number of channels in the H\&E stained brightfield images, which consist of the three channels. While the translation modules within the DD\_RoTIR model are capable of handling translations between images with three and one channels, the decision was made to compress the 3-channel H\&E brightfield images into a single channel. This approach aimed to strike a balance between the impacts of translation in both directions and maintain channel consistency in the concatenated images. Secondly, adjustments were made to the size of the inputs. The aligned images in the public dataset had original sizes of $834\times834$px$^2$, cropped from the centre of images with dimensions of $2048\times2048$px$^2$. For the testing phase, it was decided to retain the original resolutions. Consequently, subregions with dimensions of $512\times512$px$^2$ were randomly cropped from the aligned image pairs with sizes of $834\times834$px$^2$. These randomly cropped image pairs were then processed using the same procedure outlined in Section \ref{chap5:sec02} to generate the SHG dataset for testing. 

A notable difference in the training method is that models are trained within specific rotation ranges. For instance, the small rotation range is defined within the absolute range of $[0^{\circ}, 30^{\circ}]$, the medium rotation range within the absolute range of $[0^{\circ}, 45^{\circ}]$ and the large rotation range within the absolute range of $[0^{\circ}, 90^{\circ}]$. This adjustment is motivated by the observation that the SHG dataset poses a greater challenge compared to the CHO dataset. To enhance model specificity for diverse scenarios, it is essential to tailor the training focus on specific rotation ranges. Additionally, translations in both dimensions are constrained within $\pm\frac{1}{8}$ of the image size to mimic the dataset used in the CoMIR model training. The output data for the small rotation category reveals a distribution of image pairs, with a ratio of "small", "medium" and "large" corner displacements at $37\% : 34\% : 29\%$. This distribution is approximately aligned with the description in \cite{pielawski2020comir}, with the slightly smaller proportion for "small" attributed to slightly larger translations. 

The performance evaluations are conducted on four datasets, distinct from the training dataset. The evaluation dataset provides \textbf{absolute} rotation angles within specified ranges of $(0^{\circ}, 20^{\circ}]$, $(20^{\circ}, 35^{\circ}]$, $(30^{\circ}, 45^{\circ}]$ and $(45^{\circ}, 90^{\circ}]$. For each rotation range, I synthesize $768 : 384 : 576 : 720$ pairs of dual-domain images. (The ratio of numbers tests for absolute rotation angles that are lower than $45^{\circ}$ equals the ratio of the lengths of the angle ranges, maintaining the ratio of $2:1:1.5$ to ensure a balanced angle distribution). The thresholds for matrix calculation were set as 1.5 for both matching hierarchies.

\subsubsection{Testing Results}
\label{chap5:sec04sub0303}

Four DD\_RoTIR models are selected for testing on the SHG dataset. Table \ref{tab:shgconfig} outlines the configurations of the DD\_RoTIR models, with differences occurring only in the feature extraction backbone networks and rotation angle ranges. Results are presented in Table \ref{tab:shgresult}. Once again, comparing the results of DD\_RoTIR\_4M and DD\_RoTIR\_8M models, the feature extraction module with Cyclic group $C_8$ and 128 hidden layers yielded better results in all evaluation metrics. However, the improvement is not as significant as observed for the CHO dataset. 

\begin{table}[htbp]
    \centering
    \caption{Configurations of DD\_RoTIR Modes}
    \setlength{\tabcolsep}{4mm}{
    \begin{tabular}{c c c c c}
    \toprule
      Model Name   & \makecell[c]{Cyclic \\ Group} & \makecell[c]{Hidden \\ Layers} & \makecell[c]{Training \\ Range} & Size \\
    \midrule
    DD\_RoTIR\_4M  & 4 &  64 & $[0^{\circ}, 45^{\circ}]$ & 114MB \\
    DD\_RoTIR\_8L  & 8 & 128 & $[0^{\circ}, 90^{\circ}]$ &  \\
    DD\_RoTIR\_8M  & 8 & 128 & $[0^{\circ}, 45^{\circ}]$ & 139MB \\
    DD\_RoTIR\_8S  & 8 & 128 & $[0^{\circ}, 30^{\circ}]$ & \\
    \bottomrule
    \end{tabular}
    }
    \label{tab:shgconfig}
\end{table}

\begin{figure}[t!]
	\centering
	\includegraphics[width=0.9\textwidth]{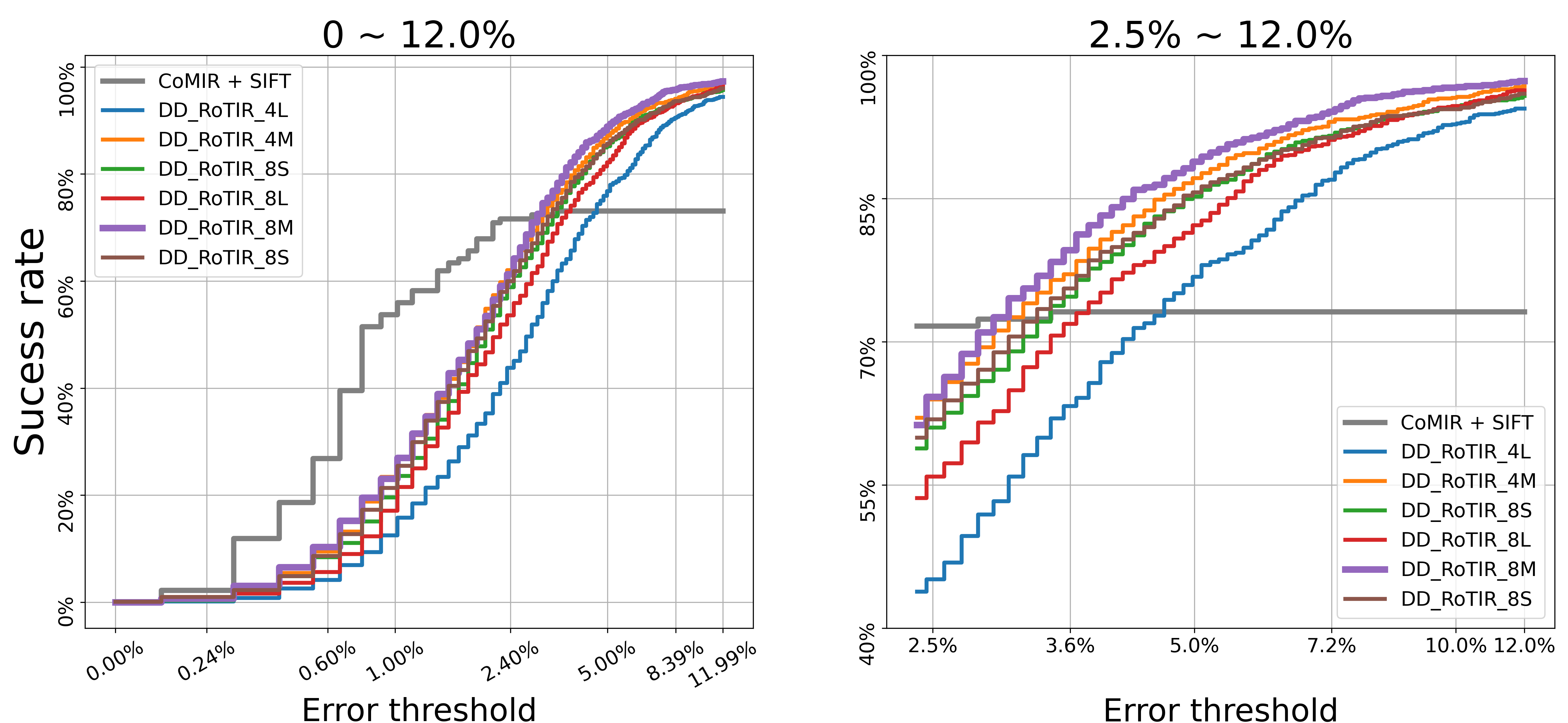}
    \caption{Relationship between error thresholds and success rates for DD\_RoTIR models with CoMIR+SIFT model as reference. All DD\_RoTIR models exhibit superior performance compared to the CoMIR+SIFT model when the error threshold reaches 5.0\%.}
	\label{fig:errsucrelat}
\end{figure}

Initially, the evaluation and result comparison was focused on the absolute rotation range of $[0^{\circ}, 30^{\circ}]$ which aligned with the CoMIR experiment. Regarding rotation detection, the DD\_RoTIR\_8M model achieved the best result in all evaluation metrics, whose average rotation difference is $1.71^{\circ}$ and corner displacement is 2.83\% relative to the image size. While the performance is slightly lower compared to the results from the CHO dataset, it remains acceptable. Given that the models share the same configuration, it is evident that the experiments on the SHG dataset pose more significant challenges. The success rates for errors of less than 5\% are remarkably higher than for DD\_RoTIR models compared to the CoMIR models with SIFT (73.1\%). However, a crucial observation is that the best success rate for errors less than 1\% (27.3\%) is only half of that achieved by the CoMIR model with SIFT (56.0\%). The relationship between the error thresholds and the success rates for DD\_RoTIR models, along with a comparison to the CoMRIR with SIFT model, are depicted in Figure \ref{fig:errsucrelat}. While the DD\_RoTIR\_8M models outperform the CoMIR+SIFT models when the error threshold exceeds 3\%, the substantial gap in success rates for errors less than 1\% needs further analysis.

\begin{table}[htbp]
    \centering
    \caption{Experiment Results on SHG dataset}
    \begin{threeparttable}
    \setlength{\tabcolsep}{6mm}{
        \begin{tabular}{| c | c | c | c | c | c |}
        \hline
        \multirow{3}*{\makecell[c]{Testing\tnote{1} \\ Dataset}} & \multirow{3}*{Metric} & \multicolumn{4}{c|}{DD\_RoTIR}  \\
        \cline{3-6}
         & & \multirow{2}*{\makecell[c]{4M\\$ 0^{\circ} \rightarrow 45^{\circ}$}} & \multirow{2}*{\makecell[c]{8L\\$ 0^{\circ} \rightarrow 90^{\circ}$}} & \multirow{2}*{\makecell[c]{8M\\$ 0^{\circ} \rightarrow 45^{\circ}$}} & \multirow{2}*{\makecell[c]{8S\\$ 0^{\circ} \rightarrow 30^{\circ}$\tnote{2}}}  \\
          & & & & & \\ 
        \hline\hline
        \multirow{4}*{$ 0^{\circ} \rightarrow 20^{\circ}$} & Angle & 1.65$^{\circ}$ & 1.87$^{\circ}$ & 1.49$^{\circ}$ & 1.76$^{\circ}$  \\ 
        \cline{2-6}
         & Corner & 2.76\% & 3.30\% & 2.55\% & 2.77\%   \\  
         \cline{2-6}
         &  < 1\% & 27.2\% & 21.1\%  & 29.0\%  & 27.1\%    \\ 
         \cline{2-6}
         &  < 5\% & 88.7\% & 84.5\%  & \textbf{90.8\%}  & 88.5\%    \\  
        \hline\hline
        \multirow{4}*{$ 0^{\circ} \rightarrow 30^{\circ}$} & Angle & 1.87$^{\circ}$ & 1.90$^{\circ}$ & 1.71$^{\circ}$ & 2.16$^{\circ}$   \\ 
        \cline{2-6}
         & Corner & 3.04\% & 3.30\% & 2.83\% & 3.29\%  \\  
         \cline{2-6}
         &  < 1\% & 26.0\% & 21.2\%  & 27.3\%  & 24.2\%   \\ 
         \cline{2-6}
         &  < 5\% & 87.3\% & 83.6\%  & \textbf{89.0\%}  & 85.4\%   \\  
        \hline\hline
        \multirow{4}*{$ 30^{\circ} \rightarrow 45^{\circ}$} & Angle & 5.09$^{\circ}$ & 2.25$^{\circ}$ & 4.08$^{\circ}$ & 8.21$^{\circ}$   \\ 
        \cline{2-6}
         & Corner & 6.85\% & 3.60\% & 5.73\% & 10.69\%  \\  
         \cline{2-6}
         &  < 1\% & 16.0\% & 22.7\%  & 18.9\%  & 9.4\%    \\ 
         \cline{2-6}
         &  < 5\% & 70.0\% & \textbf{82.6\%}  & 74.7\%  & 57.6\%    \\  
        \hline\hline
        \multirow{4}*{$ 45^{\circ} \rightarrow 90^{\circ}$} & Angle & 51.36$^{\circ}$ & 6.15$^{\circ}$ & 50.45$^{\circ}$ & 57.81$^{\circ}$   \\ 
        \cline{2-6}
         & Corner & 59.21\% & 8.04\% & 58.39\% & 66.81\%   \\  
         \cline{2-6}
         &  < 1\% & 1.4\% & 19.3\%  & 1.0\%  & 0.3\%    \\ 
         \cline{2-6}
         &  < 5\% & 14.0\% & \textbf{76.7}\%  & 14.9\%  & 6.1\%    \\  
        \hline
        \end{tabular}
    }
    \begin{tablenotes}
        \footnotesize
        \item[1] Absolute rotation angle range of the testing dataset.
        \item[2] Absolute rotation angle range of the training dataset for each model.
    \end{tablenotes}
    \end{threeparttable}
    \label{tab:shgresult}
\end{table}

\begin{figure}[t!]
\centering
\begin{minipage}{0.85\textwidth}
    \centering
    \includegraphics[width=\linewidth]{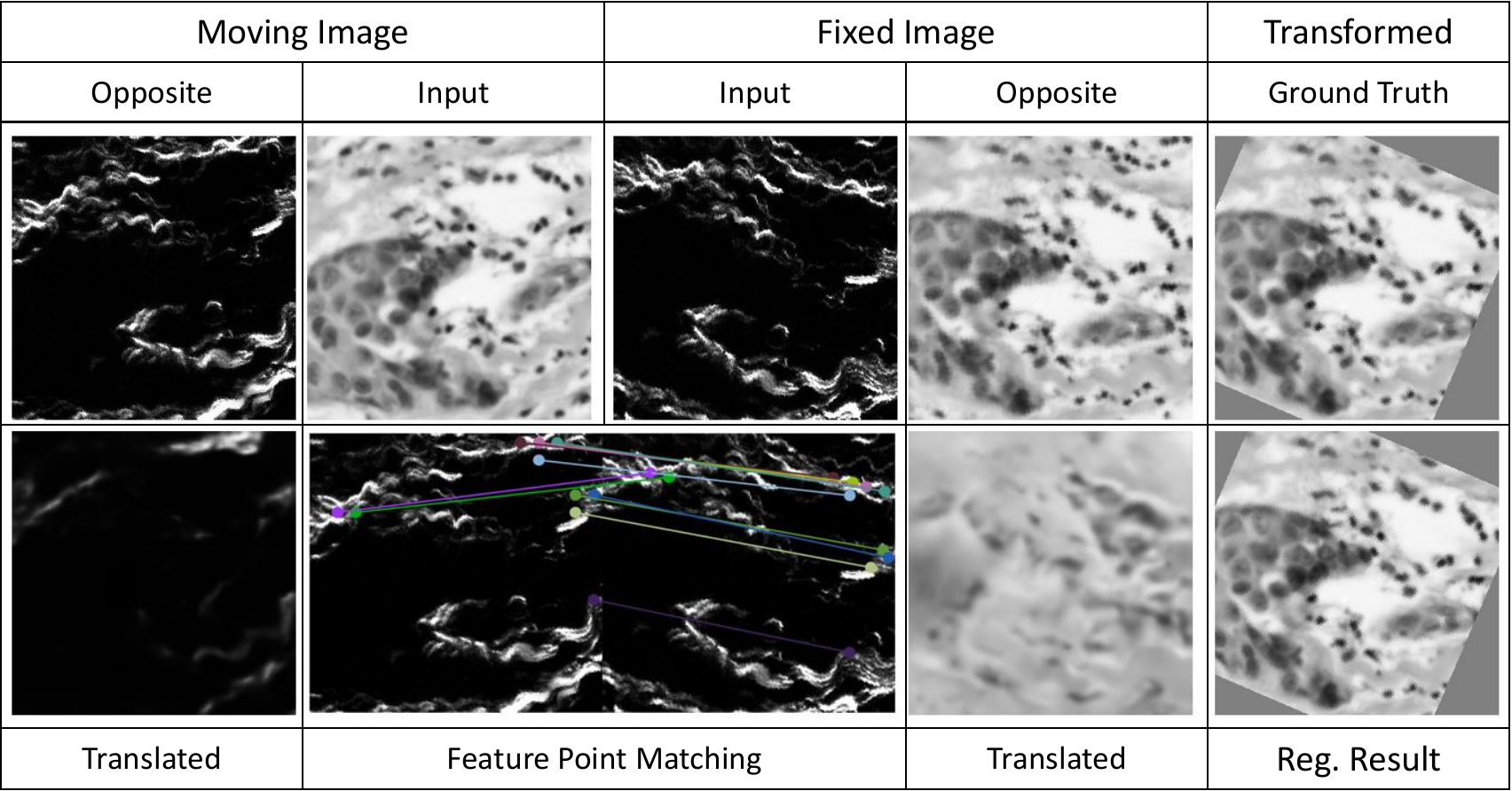}
\end{minipage}
\\ \vspace{0.1cm}
    
\begin{minipage}{0.85\textwidth}
    \centering
    \includegraphics[width=\linewidth]{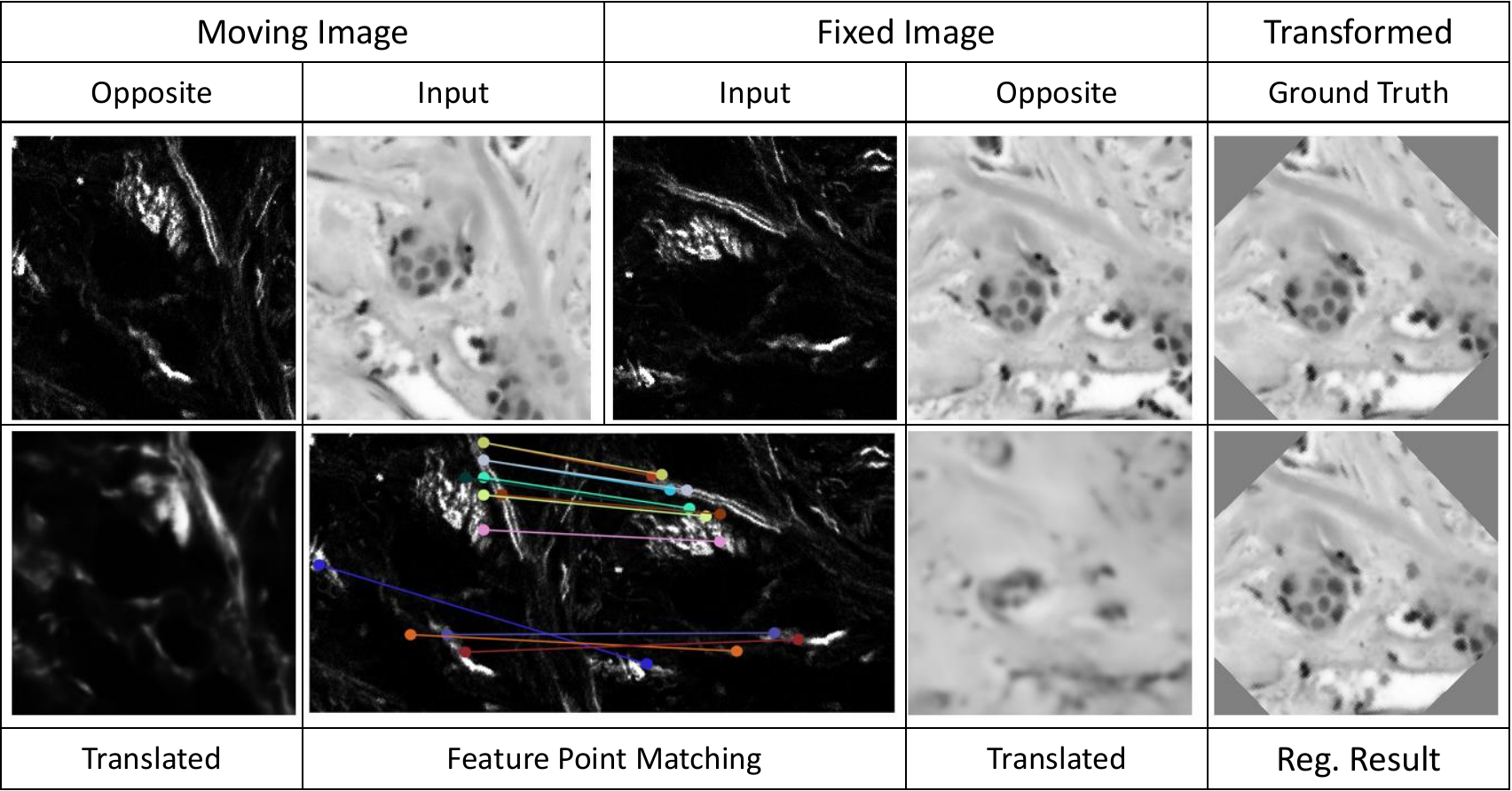}
\end{minipage}

\caption{Results by DD\_RoTIR\_8M model, trained on a dataset with absolute rotation difference within $[0^{\circ}, 45^{\circ}]$. Ground-truth images from the opposite modalities are presented beside the input images. Translated images are the outputs of translation modules within the model. The transformation of ground truth and registration results are shown on the left.}
\label{fig:8mresult}
\end{figure}

\begin{figure}[t!]
\centering
\begin{minipage}{0.85\textwidth}
    \centering
    \includegraphics[width=\textwidth]{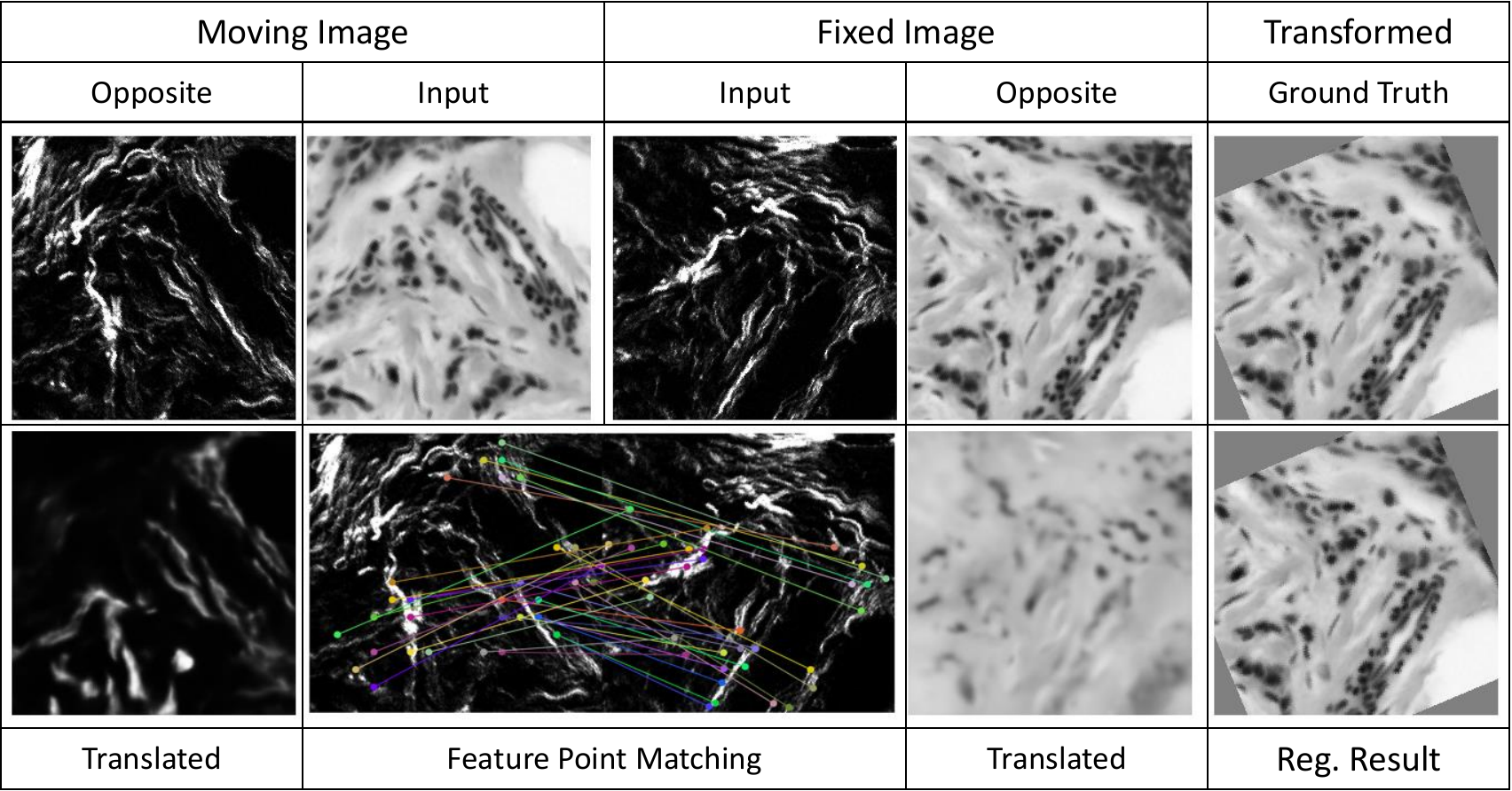}
\end{minipage}
\\ \vspace{0.1cm}
    
\begin{minipage}{0.85\textwidth}
    \centering
    \includegraphics[width=\textwidth]{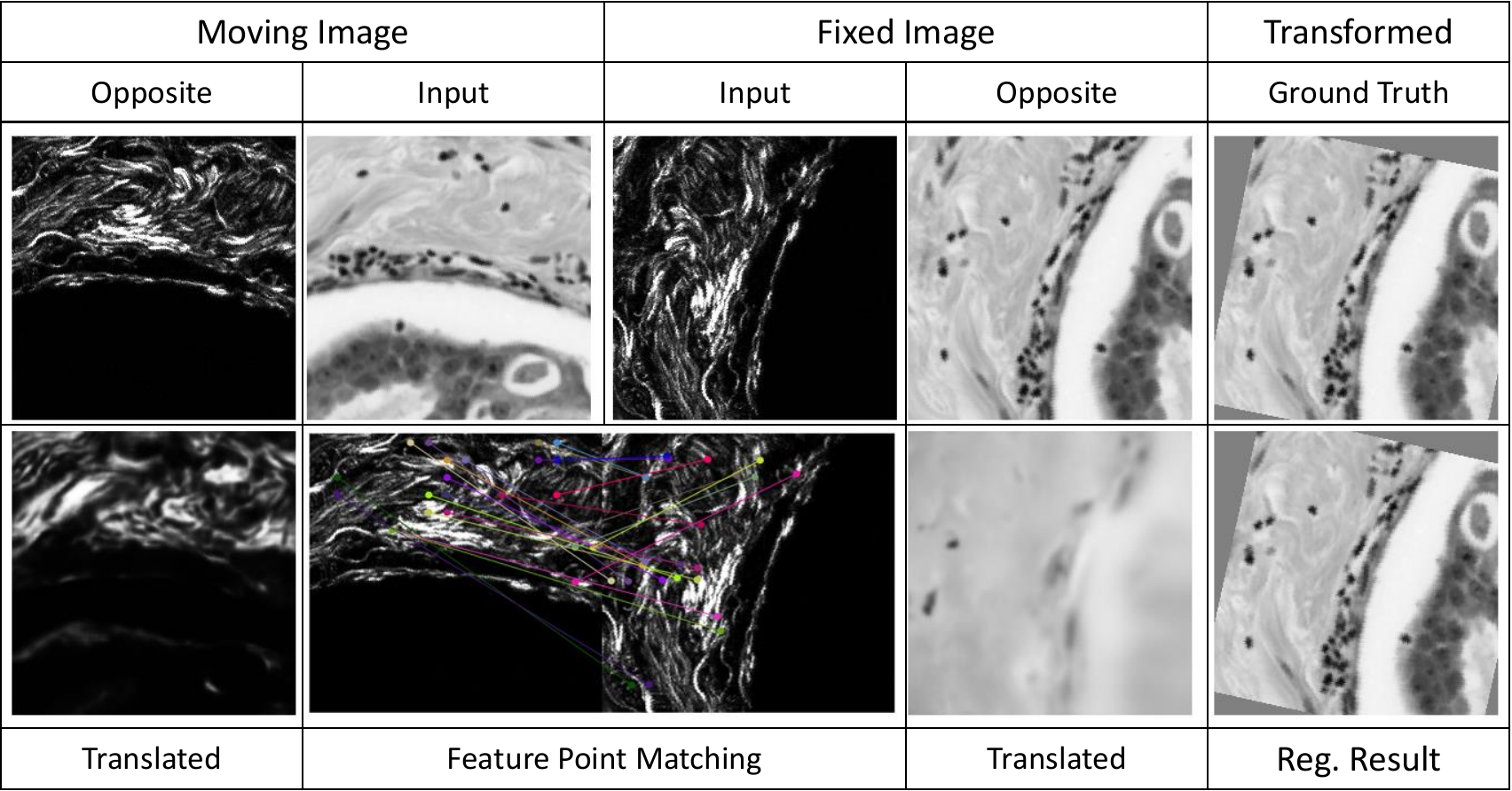}
\end{minipage}

\caption{Results by DD\_RoTIR\_8L model, model trained on dataset with absolute rotation difference within $[0^{\circ}, 90^{\circ}]$. The layouts are the same as those shown in Figure \ref{fig:8mresult}}
\label{fig:8lresult}
\end{figure}

\subsubsection{Robustness to Rotations}
\label{chap5:sec04sub0304}

\begin{figure}[t!]
	\centering
	\includegraphics[width=\textwidth]{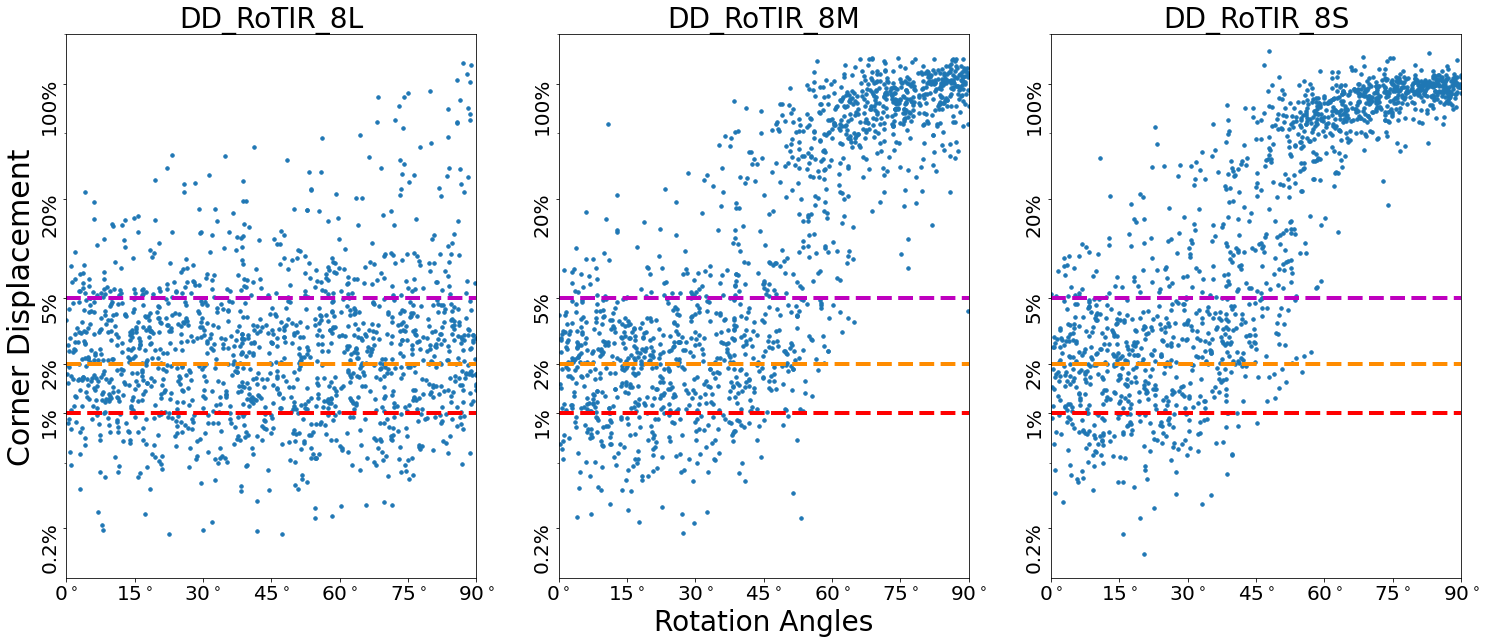}
    \caption{Relation between rotation angles and registration results. The dashed lines represent the ratios of corner displacements to the side length of the images, with 1\% (red), 2\% (orange), and 5\% (purple).}
	\label{fig:rotrelation}
\end{figure}

The rotation angles in the training dataset exert a considerable influence on the performance of the DD\_RoTIR models. As shown in table \ref{tab:shgresult}, the DD\_RoTIR\_8M model exhibits a significant advantage in the registration performance within the absolute rotation range of $[0^{\circ}, 30^{\circ}]$ compared to the DD\_RoTIR\_8L. This observation underscores the positive impact of a broader training range on registration accuracy. However, it is essential to note that a slightly wider range may not universally enhance performance for every specific rotation range. For instance, the DD\_RoTIR\_8S model is not as competitive as the DD\_RoTIR\_8M model within the absolute rotation range of $[0^{\circ}, 20^{\circ}]$ emphasizing the need for ample variety in the training dataset. Figure \ref{fig:rotrelation} visually illustrates the relationships between the ground truth of rotation differences and registration performance (represented by corner displacement) for the DD\_RoTIR\_8L, DD\_RoTIR\_8M and DD\_RoTIR\_8S models.

On the other hand, results on the rotation range that exceed the corresponding training range demonstrate immediate failure. Combining the insights from Figure \ref{fig:rotrelation} and Table \ref{tab:shgresult}, it becomes evident that the performance of DD\_RoTIR\_8M declines when applied to absolute rotation ranges of $[45^{\circ}, 90^{\circ}]$ and DD\_RoTIR\_8S to absolute rotation ranges of $[30^{\circ}, 90^{\circ}]$. However, the DD\_RoTIR\_8L model exhibits robustness across the entire rotation range, although its performance decreases with the expansion of the rotation range. The distribution of the performance remains consistent for the DD\_RoTIR\_8L model, achieving a success rate of less than 5\% error to be 76.7\%, surpassing the performance of CoMIR with SIFT on absolute rotation range of $[0^{\circ}, 30^{\circ}]$ (73.1\%). In practice applications, the choice of the DD\_RoTIR model can be tailored based on specific requirements for different scenarios.

\subsubsection{Feature Point Matching Results}
\label{chap5:sec04sub0305}

We delved into the relationship between the feature-point matching and the registration result. Ideally, two matched pairs of feature points would suffice for rigid transformation. However, in practice, errors in the detected locations of matched feature points are critical for accurate transformation matrix calculation. To mitigate the impact, more matched feature points are required. This underscores the rationale behind reducing the threshold for the fine-matching hierarchy to 1.5. Nonetheless, there exists a paradox, as a lower threshold introduces more false positive elements into the system.

\begin{figure}[t!]
	\centering
	\includegraphics[width=0.8\textwidth]{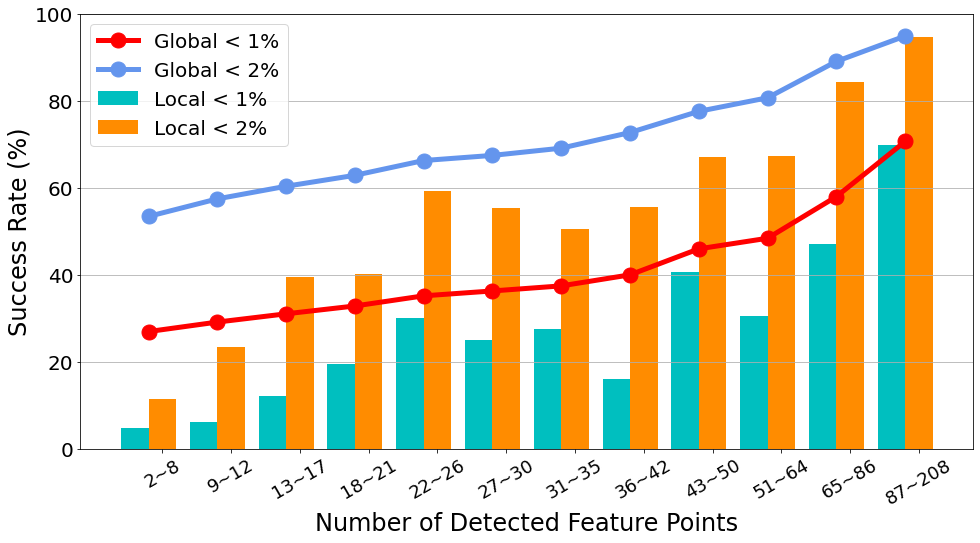}
    \caption{Relationship between registration performance and the number of detected feature points for DD\_RoTIR\_8M model. Registration cases with ground-truth rotation differences less than $30^{\circ}$ are evenly split into 12 groups based on the number of detected feature points. The bar graphs show the success rates of less 1\% and 2\% error for each group. The line charts show the respective success rates higher than the lower limit of each group.}
	\label{fig:featurenum}
\end{figure}

The experiment recorded the numbers of matching pairs for analysis. Figure \ref{fig:featurenum} illustrates the relationship between the registration performance, measured by corner displacement errors, on the dataset with an absolute rotation difference of $[0^{\circ}, 30^{\circ}]$, and the numbers of matching points detected by the DD\_RoTIR\_8M. The individual cases are categorized into 12 groups based on the number of detected feature points with equal frequency. It is evident that, as the number of detected points increases, the success ratios increase concurrently. For detected numbers higher than 13, the success rate for less than 2\% errors exceeds 60\%. Moreover, for detected numbers higher than 51, the success rate for less than 1\% errors will approach or surpass the performance of the CoMIR\_SIFT model. 

\begin{figure}[t!]
	\centering
	\includegraphics[width=0.75\textwidth]{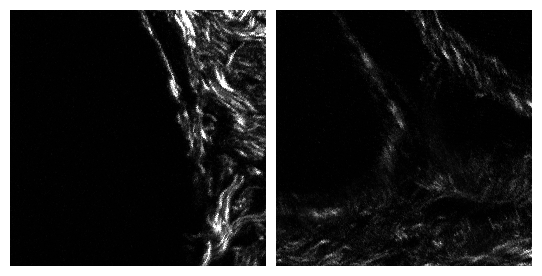}
    \caption{Examples of extreme cases in SHG images.}
	\label{fig:extrem}
\end{figure}

On the other hand, the disparity comes from the difference in the training dataset. The original data's brightfield images possess three channels, while in the experiment, I consolidated the brightfield images into a single channel. Additionally, the size of the training images for the DD\_RoTIR models ($256\times256$px$^2$) was considerably smaller than that for the CoMIR models ($834\times834$px$^2$), representing only 9.4\% of the original area. Therefore, the training images contained only $1/10$ of the information present in the original CoMIR dataset. Moreover, the intensity distributions in SHG images exhibited significant variations, with some regions containing mostly empty areas. This lack of information poses a challenge in providing sufficient matching feature points for accurate transformation matrix calculations. Figure \ref{fig:extrem} displays examples of the input SHG images. It is evident that the asymmetric intensity distributions in SHG images and the presence of defects contribute to a reduction in the information available for feature matching. This intrinsic limitation poses a challenge for achieving high-precision registration. An intuitive deduction is that by detecting and matching more feature points, the accuracy of the DD\_RoTIR\_8M model should improve, based on the statistics presented in Figure \ref{fig:featurenum}.

\section{Discussions and Future Work}
\label{chap5:sec05}

In this chapter, a novel deep learning-based model for the registration of multimodal biological microscopy images, known as DD\_RoTIR, was presented. This work draws inspiration from the success of previous approaches in image translation and mono-modal image registration. The design incorporates state-of-the-art neural networks, combining the advantages of U-Net, conditional GANs, Siamese networks, Transformers and equivariant steerable CNNs. Additionally, the concept of the artefact dataset for semi-supervised learning is introduced in the training process. The model successfully overcomes the challenges posed by multiple types of microscopy images, demonstrating effective image registration for two distinct biological image datasets. This model holds potential as a valuable tool in bio-cellular research, contributing to advanced analysis in the field.

The DD\_RoTIR models challenge the previous conclusion from the CoMIR projects that GAN-based models are unfeasible for multimodal image registration. The Siamese network-based approach for feature extraction and description unveils latent factors from pairs of real and pseudo images, which prove suitable for similarity computation. While the CoMIR model, based on contrastive representation learning, holds a nonnegligible advantage in modality information exchange, the latent space it creates is conducive to traditional image-matching methods, demonstrating superiority in precise registration but low robustness, especially in rotation, which is a common disadvantage for traditional methods. On the contrary, the DD\_RoTIR model, with multiple matching hierarchies, exhibits greater power in feature matching. Beginning with a small specificity but a high versatility approach, it achieves competent results. Each method has its own set of cons and pros, paving the way for future improvements in diverse directions.

Two primary avenues can significantly enhance the performance of the multimodal image registration task. Firstly, there is a need to refine algorithms facilitating information exchanges between modalities. While GAN-based synthetic images have proven effective, it is crucial to recognise that algorithms such as contrastive representation learning offer advantages over GAN in specific scenarios. This enhancement can streamline the feature extraction and description process. Secondly, improvements in network design for feature comparison, as demonstrated in the progression from RoTIR to DD\_RoTIR, remain an ongoing area for refinement. However, there is still space for improvement. DD\_RoTIR, same as RoTIR, employs the centroids of the sub-regions as matching candidates, sacrificing the specificity of points of interest. This strategy hinders precise location prediction on fixed image planes, leading to limitations in achieving extreme accuracy. The implementation of automatic and efficient filtration of points of interest could prove beneficial. Besides, as the computational cost of the matching algorithm escalates with increasing image size, finding a balance for large images poses a challenge. Addressing these aspects is crucial for future advancements in image registration.





\section*{Acknowledgments}
\label{sec:acknowledgments}

This work was carried out using the BlueCrystal Phase 4 facility of the Advanced Computing Research Centre, University of Bristol (\url{https://www.bristol.ac.uk/acrc/high-performance-computing/}). We thank the technical staff from the Wolfson Bioimaging Facility for their expert support. CHO image dataset can be assessed at \url{https://doi.org/10.5523/bris.2w8trsx55b0k22qez7uycg9sf4}

\bibliographystyle{plain} 
\bibliography{references}  

\clearpage
\section*{Supporting Information}

\renewcommand{\thefigure}{\thesupfigure}
\renewcommand{\theequation}{\thesupequation}

\setcounter{supsection}{0}
\setcounter{supfigure}{0}
\setcounter{supequation}{0}

\refstepcounter{supsection}
\section*{\thesupsection\ Affine Transformation Matrices}
\addcontentsline{toc}{section}{\thesupsection\ Supplementary Section}
\label{chap3:sec02sub02}

Affine transformation serves as a fundamental component in image registration, particularly for processes that do not require deformation. In essence, an affine transformation is a function that maps an affine space onto itself while preserving the dimension of any affine subspaces. This encompasses operations such as translation, rotation, scaling, shearing, etc. or arbitrary compositions of them in various orders \cite{gonzales2008digital}. It can be represented either as a matrix or as the product of a series of matrices. For 2D transformations, an affine matrix has a shape of $3 \times 3$ configuration and can be decomposed into seven parameters, one for rotation, and two for each translation, scaling and shearing in different dimensions \cite{chen2021learning, de2019deep}. In certain studies, affine transformation matrices are regressed directly without specifying particular parameters \cite{jaderberg2015spatial}. However, in my approach, I opted to extract the affine transformation parameters individually and subsequently construct the final matrices. Throughout the parameter learning processes, each parameter was simultaneously and independently learned under supervision. The parameters in my work include translation, rotation angles and scaling.

It is important to distinguish the affine transformation matrix from the coordinate transformation matrix. The affine transformation operates on the input images, generating the transformed outputs. Conversely, the coordinate transformation matrix calculates the new location for each pixel transforming. While these matrices share a relationship, they are not identical. In the PyTorch package, the \texttt{affine\_grid} and \texttt{grid\_sample} functions are provided for affine transformation conjointly. The \texttt{affine\_grid} function constructs a 4D tensor, representing a 2D grid of normalized coordinates after the affine transformation. I investigated the relationship between these two concepts and delved into the conversion laws connecting the affine transformation matrix and the coordinate transformation matrix. These conversion laws were employed during the ground-truth label calculation in the dataset synthesis and the computation of the affine transformation matrix computation in the final. 

The derivation process of the conversion law is described in Appendix \ref{App:sec01}. Examples of an affine transformation matrix and a coordinate transformation matrix are illustrated in Equation \ref{affinematrix} and Equation \ref{coordmatrix}, respectively. In these equations, $s$ represents the scale, $\theta$ denotes the rotation angle and $t_{(*, *)}$ signifies the parameter in the matrices controlling the translations. The subscripts $a$ and $c$ serve to distinguish between the types of matrices, with $a$ representing affine transformation and $c$ representing coordinate transformation. Additionally, $x$ and $y$ differentiate the dimensions. The conversion laws governing translation parameters between the matrices are outlined in Equation \ref{image2coord} and Equation \ref{coord2image}. 

\refstepcounter{supequation}
\begin{equation}\label{affinematrix}
    \mathbf{M}_{affine} = \begin{bmatrix}
    \frac{1}{s}\cdot\cos{\theta} & \frac{1}{s}\cdot\sin{\theta} & t_{a,x} \\
    -\frac{1}{s}\cdot\sin{\theta} & \frac{1}{s}\cdot\cos{\theta} & t_{a,y} \\
    0 & 0 & 1
    \end{bmatrix}
\end{equation}

\refstepcounter{supequation}
\begin{equation}\label{coordmatrix}
    \mathbf{M}_{coord} = \begin{bmatrix}
    s\cdot\cos{\theta} & s\cdot\sin{\theta} & t_{c,x} \\
    -s\cdot\sin{\theta} & s\cdot\cos{\theta} & t_{c,y} \\
    0 & 0 & 1
    \end{bmatrix}
\end{equation}

\refstepcounter{supequation}
\begin{align}\begin{aligned}\label{image2coord}
    t_{c,x} &= \left\{-\frac{L-1}{2} \cdot \Big( \cos{\theta} + \sin{\theta}\Big) - \frac{L}{2} \cdot \Big(t_{a,y} \cdot \cos{\theta} + t_{a,x} \cdot \sin{\theta}\Big) + \frac{L-1}{2}\right\} \cdot s + \frac{L-1}{2} \cdot (1-s) \\
    t_{c,y} &= \left\{-\frac{L-1}{2} \cdot \Big( \cos{\theta} - \sin{\theta}\Big) + \frac{L}{2} \cdot \Big(t_{a,y} \cdot \sin{\theta} - t_{a,x} \cdot \cos{\theta}\Big) + \frac{L-1}{2}\right\} \cdot s + \frac{L-1}{2} \cdot (1-s)
\end{aligned}\end{align}

\refstepcounter{supequation}
\begin{align}\begin{aligned}\label{coord2image}
    t_{a,x} = &\quad\frac{L-1}{L} \cdot \Big( \cos{\theta} + \sin{\theta} -1 \Big)\\
    &- \frac{2}{L} \cdot \bigg[\frac{1}{s} \cdot \Big(t_{c,x} - \frac{L-1}{2} \cdot (1-s)\Big) \cdot \sin{\theta} + \frac{1}{s} \cdot \Big(t_{c,y} - \frac{L-1}{2} \cdot (1-s)\Big) \cdot \cos{\theta} \bigg] \\
    t_{a,y} = &\quad\frac{L-1}{L} \cdot \Big( \cos{\theta} - \sin{\theta} -1 \Big)\\
    &- \frac{2}{L} \cdot \bigg[\frac{1}{s} \cdot \Big(t_{c,x} - \frac{L-1}{2} \cdot (1-s)\Big) \cdot \cos{\theta} - \frac{1}{s} \cdot \Big(t_{c,y} - \frac{L-1}{2} \cdot (1-s)\Big) \cdot \sin{\theta} \bigg] 
\end{aligned}\end{align}

\refstepcounter{supsection}
\section*{\thesupsection\ Transformation Matrix Conversion}
\addcontentsline{toc}{section}{\thesupsection\ Supplementary Section}
\label{App:sec01}

There are two ways to implement the affine transformation in the \texttt{PyTorch} and \texttt{Torchvision} libraries within a \texttt{Python} environment. The first method involves utilizing \texttt{affine} \footnote{\texttt{torchvision.transforms.functional.affine}} function in \texttt{Torchvision}. This function accepts specific transformation parameters, such as including \texttt{angle} (for rotation), \texttt{translate} (for translation) and \texttt{scale}. Unlike the second method, \texttt{affine} function does not involve affine matrices; instead, it directly yields the transformed image. This approach was employed in the experiment when constructing the artificial dataset. Two sets of independent parameters for transformation were randomly selected respectively, creating the moving and fixed images using the \texttt{affine} function. Subsequently, the matching map generation and affine matrices were determined using these two sets of parameters. The second method utilizes a combination of \texttt{affine\_grid} \footnote{\texttt{torch.nn.functional.affine\_grid}} and \texttt{grid\_sample} \footnote{\texttt{torch.nn.functional.grid\_sample}} functions. The \texttt{affine\_grid} function converts input affine matrices into flow fields (sampling grid), while \texttt{grid\_sample} takes the images and the grid flows as input and outputs the transformed images. This approach was adopted to transform the moving images after obtaining the affine matrices, specifically for presentation and evaluation purposes. 

Matrices representing affine transformation and coordinate transformation both have sizes of ($3\times3$).
The coordinate transformation is essential for determining the relationship between the coordinates of corresponding feature points on both the moving and fixed images. Therefore, the matrices for coordinate transformation differ from those for affine transformation. In this section, I first outline the process of constructing the affine transformation matrices and the coordinate transformation matrices. Subsequently, I elucidate the conversion process between these two types of matrices. This conversion plays a crucial role in dataset synthesis and generating the final results.  

\begin{figure}[t!]
	\centering
	\includegraphics[width = 0.8\textwidth]{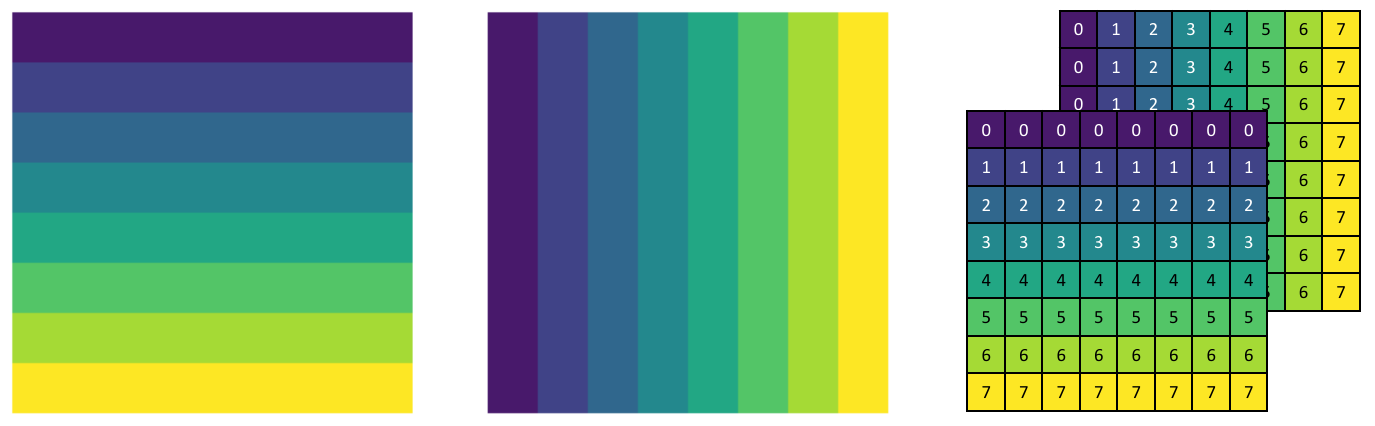}
    \caption{Mesh grids of input image. The intensity of pixels represents the coordinates in the two dimensions.}
	\label{fig:original}
\end{figure}

Figure \ref{fig:original} illustrates a pair of mesh grids representing the image intended for transformation, depicted using the viridis palette. The intensity of each pixel corresponds to its coordinates on the original image along the respective directions. These flattened mesh grids constitute the coordinate arrays of the pixels in the input images. Affine transformation of the image is accomplished through the functions \texttt{affine\_grid} and \texttt{grid\_sample}, with their outputs compared to those of the \texttt{affine} function. Coordinate transformation is facilitated via matrix multiplication with these flattened arrays. Subsequent experiments investigating the relationship between affine matrics and coordinate transformation will utilize these arrays. The flattened arrays are represented by matrices with sizes of $(2 \times L)$, where $2$ denotes the number of dimensions and $L$ is the length of the arrays. These matrices are concatenated with arrays of the same length, where all elements are $1$, forming matrices of sizes ($3\times L)$. Consequently, the multiplication order becomes $\mathbf{M}_{coord}\mathbf{A}_{coord}$.

\refstepcounter{supfigure}

\subsection*{\thesupsection.1\ Translation}
\label{App:sec01sub01}

\refstepcounter{supfigure}
\begin{figure}[t]
    \centering
    \begin{subfigure}[b]{\textwidth}
        \centering
        \includegraphics[width=0.8\textwidth]{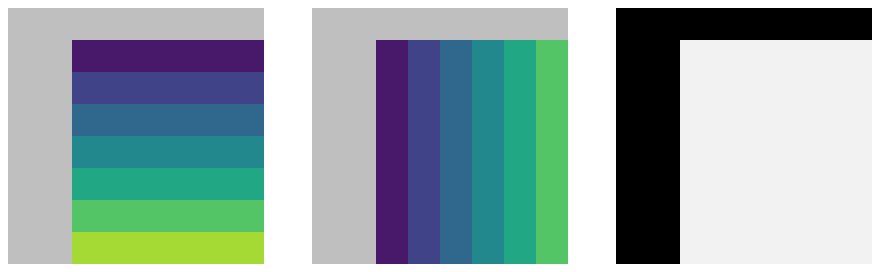}
        \caption{Outcome by \texttt{affine}.}
    \end{subfigure}
    \vspace{0.1cm} 
    \begin{subfigure}[b]{\textwidth}
        \centering
        \includegraphics[width=0.8\textwidth]{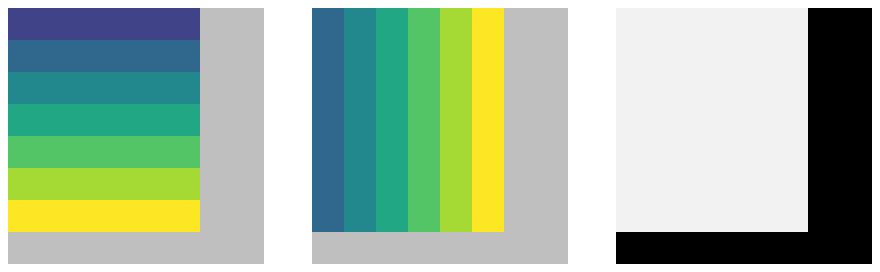}
        \caption{Affine transformation.}
    \end{subfigure}
    \vspace{0.1cm} 
    \begin{subfigure}[b]{\textwidth}
        \centering
        \includegraphics[width=0.8\textwidth]{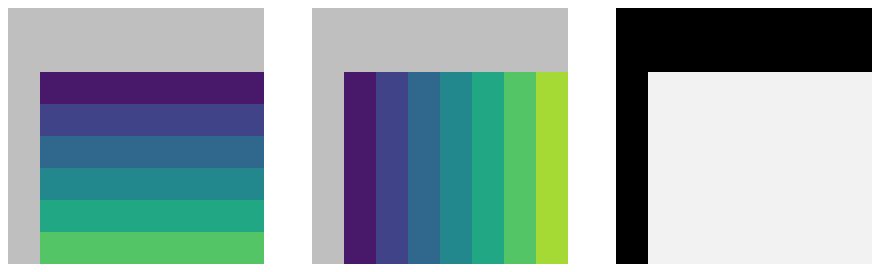}
        \caption{Coordinate transformation.}
    \end{subfigure}
    \caption{Translation outcomes by (a) function, \texttt{affine}, (b) affine transformation by Matrix \ref{app:transamat}, (c) coordinate transformation using Matrix \ref{app:transcmat}. The last columns are the masks that indicate the original components remained on the output images.}
    \label{fig:apptranslation}
    \vspace{0.1cm}  
\end{figure}

Initially, I investigated the relationship between input values for \texttt{affine} function and the resulting affine matrix. I selected the tuple $(d_x, d_y)=(2, 1)$, which were 0.5 and 0.25 of the half sizes of the images ($\ \frac {L}{2}$), as the input for the \texttt{tranlate} variable. Subsequently, $3\times3$ matrices were formed for affine and coordinate transformations. It is important to note that the elements in the affine matrices controlling translation, the first two elements in the third column, were regularized by dividing by half the size of the images. These matrices are presented in Matrix \ref{app:transamat} and \ref{app:transcmat}.

\hspace{0.05\textwidth}
\begin{minipage}{0.35\textwidth}
\centering
\refstepcounter{supequation}
\begin{equation}\label{app:transamat}
    \mathbf{M} = \begin{bmatrix}
    1 & 0 & 0.5  \\
    0 & 1 & 0.25 \\
    0 & 0 & 1
    \end{bmatrix}
\end{equation}
\end{minipage}
\hspace{0.05\textwidth}
\begin{minipage}{0.35\textwidth}
\centering
\refstepcounter{supequation}
\begin{equation}\label{app:transcmat}
    \mathbf{M} = \begin{bmatrix}
    1 & 0 & 2  \\
    0 & 1 & 1 \\
    0 & 0 & 1
    \end{bmatrix}
\end{equation}
\end{minipage}

Figure \ref{fig:apptranslation} illustrates the test results for translation. Figure \ref{fig:apptranslation}(a) presents the results obtained using the \texttt{affine} function, which aligns with the desired outcomes. The results of affine transformation implemented by Matrix \ref{app:transamat} and Matrix \ref{app:transcmat} are shown in Figures \ref{fig:apptranslation}(a) and (b), respectively. Comparing Figures \ref{fig:apptranslation}(a) and (b), I note that the elements are correctly positioned in the matrix, controlling the correct dimensions; however, the translational directions are inverted. In contrast, Figure \ref{fig:apptranslation}(c) demonstrates that when applying to coordinates, the two elements in the last row should be in reverse positions with the same signs to align with the result in Figure \ref{fig:apptranslation}(a). Finally, the correlation of the affine matrix and coordinate transformation matrix is depicted in Equation \ref{app:transconv}.

\refstepcounter{supequation}
\begin{equation}\label{app:transconv}
    \mathbf{M}_{affine} = \begin{bmatrix}
    1 & 0 & -d_x / \frac{L}{2}  \\
    0 & 1 & -d_y / \frac{L}{2} \\
    0 & 0 & 1
    \end{bmatrix} \iff \mathbf{M}_{coord} = \begin{bmatrix}
    1 & 0 & d_y  \\
    0 & 1 & d_x \\
    0 & 0 & 1
    \end{bmatrix}
\end{equation}

\subsection*{\thesupsection.2\ Rotation}
\label{App:sec01sub02}

Subsequently, I constructed the rotation matrix using the sine and cosine values of $\theta$, as shown in Matrix \ref{app:rotmat}, and repeated the operation in Section \ref{App:sec01sub01}. 

\refstepcounter{supequation}
\begin{equation}\label{app:rotmat}
    \mathbf{M} = \begin{bmatrix}
    \cos{\theta} & \sin{\theta} & 0 \\
    -\sin{\theta} & \cos{\theta} & 0 \\
    0 & 0 & 1
    \end{bmatrix}
\end{equation}

    

\refstepcounter{supfigure}
\begin{figure}[t]
    \centering
    \begin{subfigure}[b]{\textwidth}
        \centering
        \includegraphics[width=0.8\textwidth]{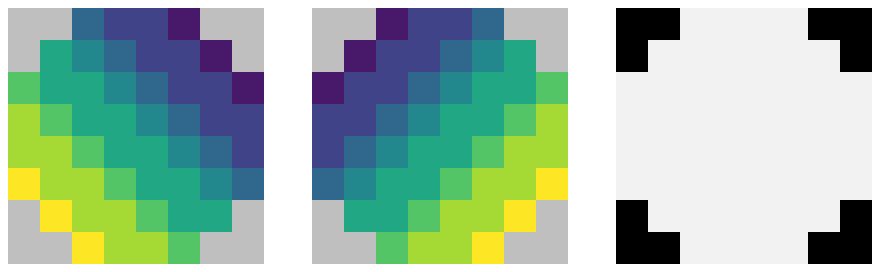}
        \caption{Outcome by \texttt{affine} \& Affine transformation.}
    \end{subfigure}
    \vspace{0.1cm} 
    \begin{subfigure}[b]{\textwidth}
        \centering
        \includegraphics[width=0.8\textwidth]{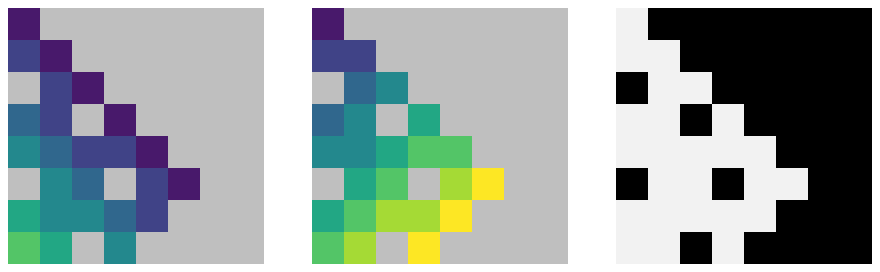}
        \caption{Coordinate transformation using affine matrix.}
    \end{subfigure}
    \caption{Rotation performance on Matrix \ref{app:rotmat}, when $\theta = 45^{\circ}$. (a) Affine transformation result. (b) Coordinate transformation using affine matrix. The images in the last columns are the masks that indicate the original components remained on the output images.}
    \label{fig:approtation}
    \vspace{0.1cm}  
\end{figure}

Figure \ref{fig:approtation} illustrates the effects of the rotation matrix when rotation angle $\theta = 45^{\circ}$. Both the results obtained using the function \texttt{affine} and the affine transformation are identical and shown in Figure \ref{fig:approtation}(a). From Figure \ref{fig:approtation}(a), I observe that the image rotates clockwise around its centroids. On the contrary, in Figure \ref{fig:approtation}(b), applying Matrix \ref{app:rotmat} to the coordinate arrays, results in a clockwise rotation but around the origin (top left of the image). (The holes in the last column of Figure \ref{fig:approtation}(b) are caused by the rounding operation, as some pixels overlap to the same destination points.) This phenomenon suggests that to achieve identical transformation outcomes, the matrix for the coordinate transformation needs to reposition the image centroid to the origin before rotation and then move it back afterwards. This operation can be implemented through matrix multiplication, as shown in Equation \ref{app:rotconv}. (Notably, the combined coordinate transformation matrix follows a backward multiplication order.)  

\refstepcounter{supequation}
\begin{align}
    \mathbf{M}_{affine} &= \begin{bmatrix}
    \cos{\theta} & \sin{\theta} & 0  \\
    -\sin{\theta} & \cos{\theta} & 0 \\
    0 & 0 & 1
    \end{bmatrix} \iff \nonumber \\
    \mathbf{M}_{coord\ } &= \begin{bmatrix}
    1 & 0 & \frac{L-1}{2}  \\
    0 & 1 & \frac{L-1}{2} \\
    0 & 0 & 1
    \end{bmatrix} \begin{bmatrix}
    \cos{\theta} & \sin{\theta} & 0  \\
    -\sin{\theta} & \cos{\theta} & 0 \\
    0 & 0 & 1
    \end{bmatrix} \begin{bmatrix}
    1 & 0 & -\frac{L-1}{2}  \\
    0 & 1 & -\frac{L-1}{2} \\
    0 & 0 & 1
    \end{bmatrix} \label{app:rotconv}\\
    &= \begin{bmatrix}
    \cos{\theta} & \sin{\theta} & \frac{L-1}{2}(1 - \cos{\theta} - \sin{\theta})  \\
    -\sin{\theta} & \cos{\theta} & \frac{L-1}{2}(1 + \cos{\theta} - \sin{\theta}) \\
    0 & 0 & 1
    \end{bmatrix} \nonumber
\end{align}

\subsection*{\thesupsection.3\ Scaling}
\label{App:sec01sub03}

\begin{figure}[t]
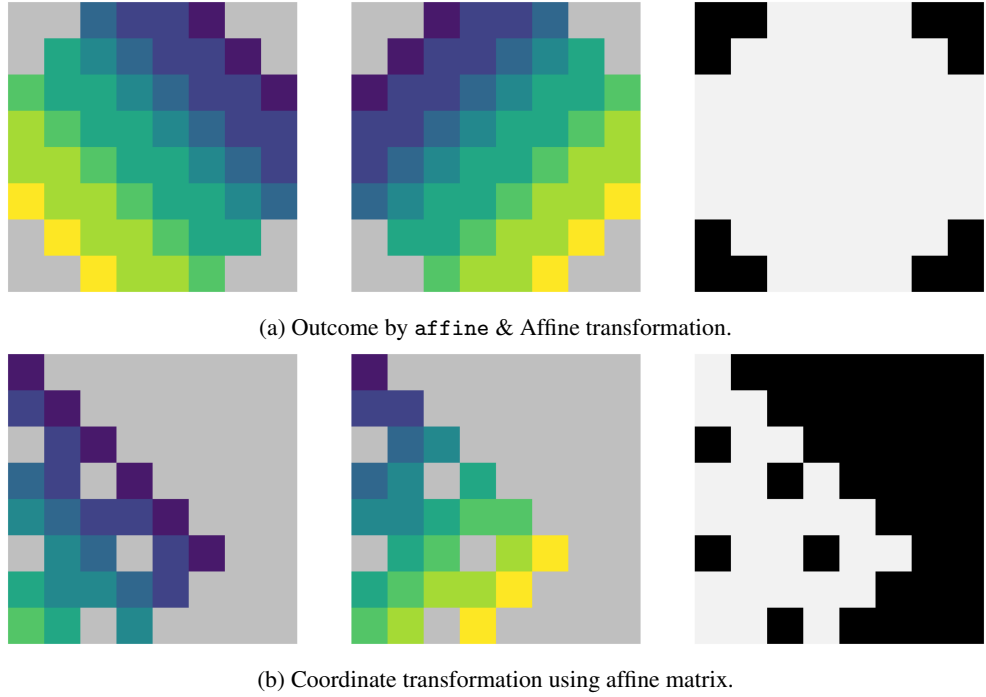

    \centering
    \begin{subfigure}[b]{\textwidth}
        \centering
        \includegraphics[width=0.8\textwidth]{app0A/rotate_w.png}
        \caption{Outcome by \texttt{affine} \& Affine transformation.}
    \end{subfigure}
    \vspace{0.1cm} 
    \begin{subfigure}[b]{\textwidth}
        \centering
        \includegraphics[width=0.8\textwidth]{app0A/rotate_f.png}
        \caption{Coordinate transformation using affine matrix.}
    \end{subfigure}
    \caption{Scaling outcomes on affine matrix, when $s = 0.5$. (a) Affine transformation result. (b) Coordinate transformation using affine matrix.}
    \label{fig:appscale}
    \vspace{0.1cm}  
\end{figure}

Following a similar analysis to that of rotation, I conducted tests using the scaling variable $s$ with the \texttt{affine} function. The result is shown in Figure \ref{fig:appscale}(a). The affine matrix is formed, shown as in Matrix \ref{app:scamat}, to achieve identical outcomes, where the first two elements on the diagonal, governing the zoom function, are the reciprocals of the scaling variable, $\frac{1}{s}$. However, upon applying Matrix \ref{app:scamat}, I observed the opposite zoom function, as shown in Figure \ref{fig:appscale}(b). Additionally, I noted that the reference point for the scaling transformation is not at the image's centroid but at the origin, consistent with the observation made in the rotation transformation described in Section \ref{App:sec01sub02}. Thus, based on these findings, I conclude that to replicate the performance of \texttt{affine} function, the coordinate matrix should first move the coordinates from the centroid to the origin, conduct the scaling transformation, and then move them back. The correlation is illustrated in Equation \ref{app:scaconv}. 

\refstepcounter{supequation}
\begin{equation}\label{app:scamat}
    \mathbf{M} = \begin{bmatrix}
    \frac{1}{s} & 0 & 0 \\
    0 & \frac{1}{s} & 0 \\
    0 & 0 & 1
    \end{bmatrix}
\end{equation}

\refstepcounter{supequation}
\begin{equation}\label{app:scaconv}
    \mathbf{M}_{affine} = \begin{bmatrix}
    \frac{1}{s} & 0 & 0  \\
    0 & \frac{1}{s} & 0 \\
    0 & 0 & 1
    \end{bmatrix} \iff \mathbf{M}_{coord} = \begin{bmatrix}
    s & 0 & \frac{L-1}{2}(1 - s)  \\
    0 & s & \frac{L-1}{2}(1 - s) \\
    0 & 0 & 1
    \end{bmatrix}
\end{equation}

    

\refstepcounter{supfigure}

\subsection*{\thesupsection.4\ Transformation Combination}
\label{App:sec01sub04}

After studying the correlation between affine and coordinate transformations for each factor individually, I can combine them to yield the composite transformation. Both transformations can be regarded as a series of individual transformations combined through matrix multiplication. Since matrix multiplication is not commutative, it is crucial to clarify the operation orders of the transformations. In the case of \texttt{affine} function, the operation order begins with rotation and scaling, with their order being interchangeable, followed by translation. Thus, when constructing the training dataset, I calculated the matrices for moving and fixed from the base image, with the target of interest located at the centre of the image, using $ \mathbf{M}_{rotate} \mathbf{M}_{scale} \mathbf{M}_{tranlate}$, (the combination of the affine matrix follows a forward multiplication order, contrasting with the coordinate transformation matrix). Subsequently, the matrix for transforming the moving image to the fixed image can be calculated using $\mathbf{M}_{movinig}^{-1} \mathbf{M}_{fixed}$. 

\subsection*{\thesupsection.5\ Conversion Laws}
\label{App:sec01sub05}

Converting between the two types of transformation is crucial in my work, as the locations of the matched points play an essential role in generating matching maps in dataset synthesis and affine matrix calculation for the final results. A given affine matrix can be directly factorized, as shown in Equation \ref{app:order}. Therefore the transformation can be regarded as a series of individual transformations in the order of translation followed by rotation and scaling. Individual terms in the affine matrix can be converted to the corresponding coordinate transformation matrix based on the Equations \ref{app:transconv}, \ref{app:rotconv} and \ref{app:scaconv}. The final conversion equations are shown in Equation \ref{affinematrix} and \ref{image2coord}. Reversely, the coordinate translation matrix can be converted to the affine transformation matrix with the conversion equations shown in Equation \ref{coordmatrix} and \ref{coord2image}.

\refstepcounter{supequation}
\begin{equation}\label{app:order}
    \mathbf{M}_{affine} = \begin{bmatrix}
    s\cdot \cos{\theta} & s\cdot \sin{\theta} & d_x  \\
    -s\cdot \sin{\theta} & s\cdot \cos{\theta} & d_y \\
    0 & 0 & 1
    \end{bmatrix} = \begin{bmatrix}
    1 & 0 & d_x  \\
    0 & 1 & d_y \\
    0 & 0 & 1
    \end{bmatrix} \begin{bmatrix}
    \cos{\theta} & \sin{\theta} & 0  \\
    -\sin{\theta} & \cos{\theta} & 0 \\
    0 & 0 & 1
    \end{bmatrix} \begin{bmatrix}
    s & 0 & 0  \\
    0 & s & 0 \\
    0 & 0 & 1
    \end{bmatrix}
\end{equation}

\refstepcounter{supsection}
\section*{\thesupsection\ Comparison of Dual-Softmax Operator and Sinkhorn Iteration Algorithm}
\addcontentsline{toc}{section}{\thesupsection\ Supplementary Section}
\label{App:sec03}

\refstepcounter{supfigure}
\begin{figure}[ph]
	\centering
	\includegraphics[width=0.85\textwidth]{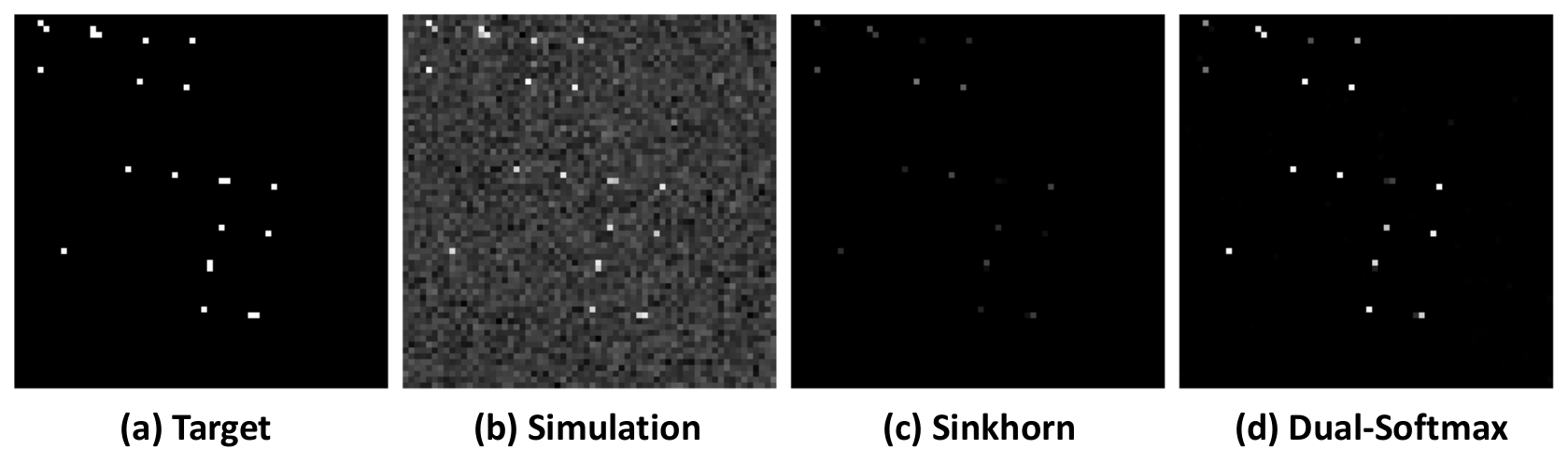}
	\caption{Comparison of confidence map generated by Sinkhorn iteration algorithm and dual-softmax operator in multi-to-multi matching situation. (a) Target confidence map, (b) Simulated unnormalized output, (c) Post-processed output by Sinkhorn iteration algorithm (d) Post-processed outputs by dual-softmax operator.}
	\label{fig:dsm_skh}
\end{figure}

The experiment aimed to compare the effect of normalization on feature-matching results and contributed to the selection between the dual-softmax operator and the Sinkhorn iteration algorithm. It was conducted on the confidence maps generated by the projection layers before normalization. Observations indicated that the maximum output of the unnormalized confidence maps typically hovered around zeros, with differences between the maximum and the minimum values ranging from 10 to 30. To simulate the scenario of one-to-multi matching, I generated matrices derived from the coarse matching maps obtained from the training dataset of CHO cell images. In these matrices, positive pixels were assigned values 10 units higher than the negative pixels, and Gaussian noise, following normal distribution, was added. 

We simultaneously applied the Sinkhorn iteration algorithm and dual-softmax operator to an identical matrix. One example result is depicted in Figure \ref{fig:dsm_skh}. Notably, the output from the Sinkhorn iteration struggled to distinguish distinct positive points from the background, with a maximum value not exceeding 0.5. On the contrary, the output from the dual-softmax operator successfully filtered out most positive points, with 15 of them scoring higher than 0.5 (12 of them higher than 0.8). 

It is worth noting that this scenario mimics an ideal case where the raw outcome closely aligns with the target. In practice applications, maps generated by the Sinkhorn algorithm often struggle to provide a feasible result for potential one-to-multi matching scenarios. Conversely, confidence maps generated by the duel-softmax operator for one-to-multi matching are more stable, which is beneficial for creating ground-truth confidence maps and aiding in loss calculation.


\end{document}